\documentclass[epj,final]{svjour}

\usepackage{epsfig}
\usepackage{amsmath}
\usepackage{amssymb}
\usepackage[dvips]{color}
\usepackage{wrapfig}
\usepackage{bm}

\newcommand{\vo}{\mbox{$V^0$}}
\newcommand{\lam}{\mbox{$\rm \Lambda$}}
\newcommand{\lamdecay}{\mbox{$\rm \Lambda \to p \pi^-$}}

\newcommand{\ko}{\mbox{$K^0_S$}}
\newcommand{\kodecay}{\mbox{$K^0_S \to \pi^+ \pi^-$}}

\newcommand{\Kdecay}{\mbox{$K^\star \to K^0_S \pi$}}

\begin{document}
\hugehead

\title{Search for the exotic $\Theta^+$ resonance in the NOMAD experiment}

\subtitle{Nomad Collaboration}

\author{
O.~Samoylov\inst{6} \and
D.~Naumov\inst{6,7} \and 
V.~Cavasinni\inst{16} \and
P.~Astier\inst{14} \and 
D.~Autiero\inst{8} \and
A.~Baldisseri\inst{18} \and
M.~Baldo-Ceolin\inst{13} \and
M.~Banner\inst{14} \and
G.~Bassompierre\inst{1} \and
K.~Benslama\inst{9}
% ...
}

\mail{naumov@nusun.jinr.ru (D.~Naumov)}

\institute{LAPP, Annecy, France \and 
Johns Hopkins Univ., Baltimore, MD, USA \and
Harvard Univ., Cambridge, MA, USA \and
Univ. of Calabria and INFN, Cosenza, Italy \and
Dortmund Univ., Dortmund, Germany \and
JINR, Dubna, Russia \and
Univ. of Florence and INFN,  Florence, Italy \and
CERN, Geneva, Switzerland \and
University of Lausanne, Lausanne, Switzerland \and
UCLA, Los Angeles, CA, USA \and
University of Melbourne, Melbourne, Australia \and
Inst. for Nuclear Research, INR Moscow, Russia \and
Univ. of Padova and INFN, Padova, Italy \and
LPNHE, Univ. of Paris VI and VII, Paris, France \and
Univ. of Pavia and INFN, Pavia, Italy \and
Univ. of Pisa and INFN, Pisa, Italy \and
Roma Tre University and INFN, Rome, Italy \and
DAPNIA, CEA Saclay, France
% ...
}

\abstract{
A search for exotic $\Theta^+$ baryon via $\Theta^+\to p+\ko$ decay mode in the NOMAD $\nu_\mu N$ data is reported. The special background generation procedure was developed. The proton identification criteria are tuned to maximize the sensitivity to the $\Theta^+$ signal as a function of $x_F$ which allows to study the $\Theta^+$ production mechanism. 
We do not observe any evidence for the $\Theta^+$ state in the NOMAD data. We provide an upper limit on $\Theta^+$ production rate at 90\% CL as $2.13\cdot 10^{-3}$ per neutrino interaction.
}

\date{Received: date / Revised version: date}

\markboth{Nomad Collaboration: Search for the exotic $\Theta^+$ resonance}
{Nomad Collaboration: Search for exotic $\Theta^+$ resonance}

\maketitle

\section{\label{sec:intro} Introduction}
In the last three  years an intense  experimental activity has been carried 
out to search for exotic baryon states with charge and flavor 
requiring a minimal valence quark configuration of four 
quarks and one antiquark (such states are often referred to as
``pentaquarks'').
Searches for exotic baryon states have a $\sim$ 30
year history, but a theoretical paper by Diakonov, Petrov and Polyakov
~\cite{Diakonov} has triggered practically all recent activity.

The LEPS Collaboration was the first to report 
the observation of the $\Theta^+$ ($uudd\bar s$) state with positive
strangeness \cite{Spring-8_1,Spring-8_2}. Then confirmations followed from 
DIANA (ITEP)~\cite{Diana}, CLAS~\cite{Clas_1,Clas_2,Clas_3,Clas_4}, 
ELSA (SAPHIR)~\cite{Saphir}, old (anti) neutrino bubble chambers data
(WA21, WA25, WA59, E180, E632) reanalyzed by ITEP physicists~\cite{Bebc}, 
HERMES~\cite{Hermes}, SVD (IHEP)~\cite{Svd}, COSY-TOF~\cite{Cosy-tof},
LHE (JINR)~\cite{lhe}, HEP ANL -- HERA (ZEUS)~\cite{Hera1,Hera2}. 
A narrow peak in the invariant mass distributions 
of $p \ko$ or $n K^+$ pairs with a mass of $\simeq 1530-1540$ MeV$/c^2$
and a width of less than 25 MeV$/c^2$ was observed 
in all these experiments with  significances  of 4-8 $\sigma$'s.
Searches for narrow pentaquark states were then performed in
almost every accelerator experiment in the world, providing evidence
or hints for a variety of pentaquark candidates: $\Theta^+$, $\Xi_{5}^{--}$,
$\Theta^{++}$, and $\Theta_c^0$.
However, after this initial flurry of positive results, negative results, 
in particular from high statistics experiments,
started to dominate the field.
As an example of a negative search we quote the HERA-B experiment at 
DESY~\cite{Hera} that observed 
neither the $\Theta^+$ resonance in the $p \ko$ invariant mass
distribution nor the $\Xi^{--}(1860)$ (another member of the antidecuplet 
of exotic baryons) decaying to $\Xi^- \pi^-$. 
Also the BES Collaboration~\cite{Bai:2004gk}
reported no $\Theta(1540)$ signal in $\psi(2S)$ and $J/\psi$ 
hadronic decays to $\ko p K^- \bar n$ and $\ko \bar p K^+ n$.
The PHENIX experiment at RHIC~\cite{Pinkenburg:2004ux} 
has seen no anti--pentaquark $\bar{\Theta}^-$ in the decay channel
$K^- \bar{n}$. Also the BABAR~\cite{Babar} and the
CDF~\cite{CDF} experiments have provided no evidence for $\Theta^+$.
Possible explanations for such a controversial experimental situation
could be ascribed to specific production mechanisms yielding pentaquarks only
for specific initial state particles.
However, the CLAS experiment at Jefferson Lab has recently reported
the results of a new analysis of photon--deuterium interactions with 
a statistics six times larger than the earlier event sample which
showed a positive result. In this new analysis no $\Theta^+$ peak was 
seen~\cite{Clas_5}. A review of the experimental evidence for and
against the existence of pentaquarks is presented in \cite{Burkert}.

This article describes a search for the lightest member of the
antidecuplet of exotic baryons, $\Theta^+$,
in the decay channel $\Theta^+\to p+\ko$ from a large sample of
neutrino interactions recorded in the NOMAD experiment at CERN.

The paper is organized as follows. In Sec.~\ref{sec:nomad-detector} we give
a brief description of the NOMAD detector, and of the NOMAD simulation
program (MC). In Sec.~\ref{sec:selection} we present the event selection 
criteria and the tools for $\ko$ and proton identification. In
Sec.~\ref{sec:selection} we describe checks of the proton identification
procedure and discuss the expected invariant mass resolution of 
$p\ko$ pairs. We describe in detail our 
procedure for determining the shape of the background distribution in
Sec.~\ref{sec:background}. Based on the background determination 
procedure and proton identification,
we then develop a strategy for a ``blind`` analysis of the 
$\Theta^+$ signal by finding the proton identification criteria
which maximize the sensitivity to the 
expected signal. This approach is presented in
Sec.~\ref{sec:theta_tools}.
In this section we present also the method to estimate the signal
significance and check the analysis chain using the observed decays
$\lam\to p+\pi^-$ and $\ko\to\pi^+\pi^-$. Finally, we ``open the box``,
i.e. examine the signal in the data. The conclusions are 
drawn in Sec.~\ref{sec:conclusion}.

\section{\label{sec:nomad-detector} The NOMAD detector}
\begin{figure*}[htb]
\begin{center}
   \mbox{
     \epsfig{file=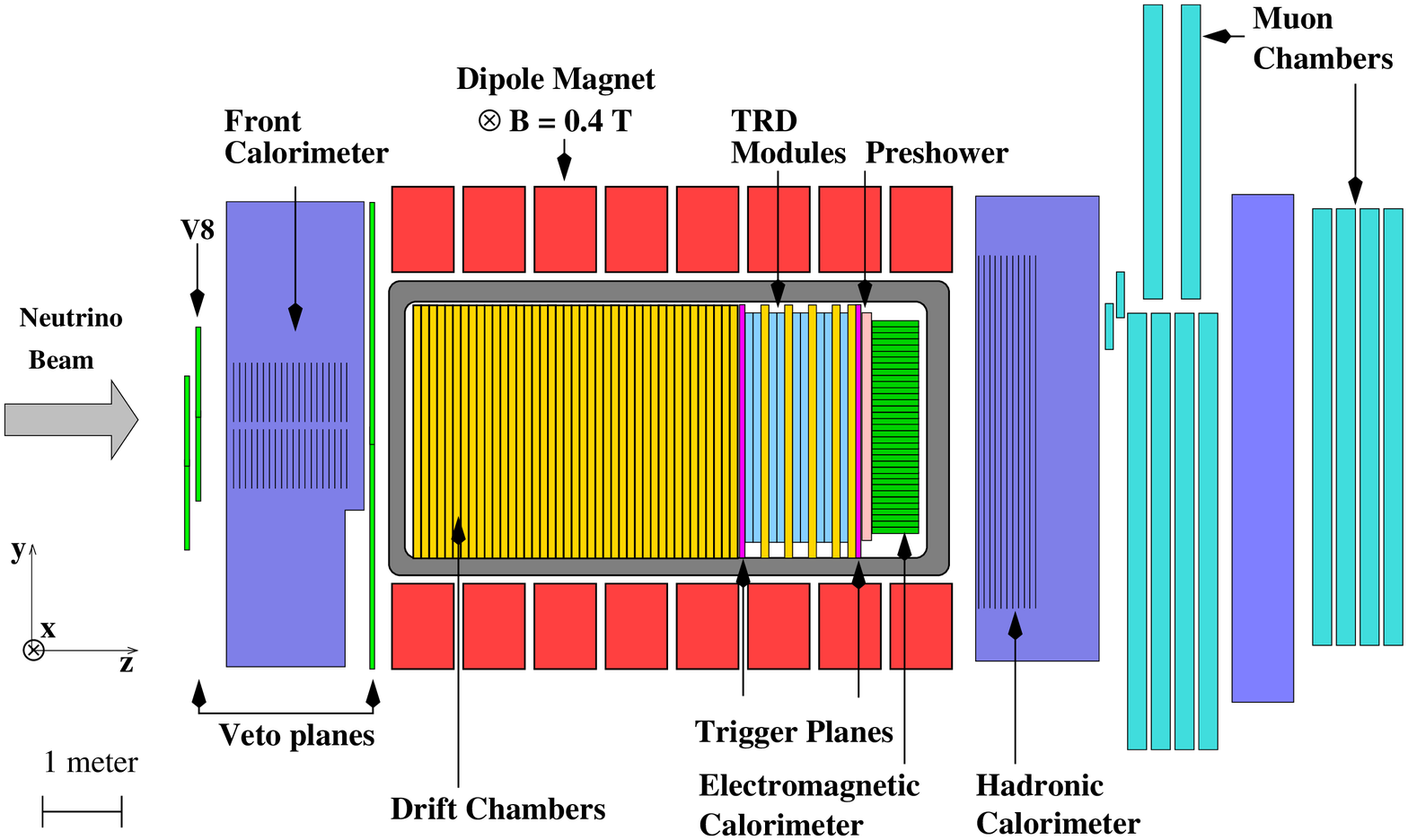,width=0.8\textwidth}}
     \caption{Side view of the NOMAD detector.}
      \label{fig:nomad_detector}
   \end{center}
\end{figure*}

The large sample of neutrino 
interactions, about 1.5 millions, measured in NOMAD  together with the good 
reconstruction quality of individual tracks, offer an excellent
opportunity to search for $\Theta^+\to p+\ko$. The
NOMAD detector ~\cite{Altegoer:1998gv} consisted of an active target of 44
drift chambers, with a total fiducial mass 
of 2.7~tons, located in a 0.4~Tesla dipole magnetic field, as shown in
Fig.~\ref{fig:nomad_detector}. 

The drift chambers~\cite{Anfreville:2001zi}, made of low $Z$ material
(mainly Carbon) served the double role of 
a nearly isoscalar target for neutrino interactions
and of the tracking medium. 
The average density of the drift chamber volume was 
0.1 $\mbox{g}/\mbox{cm}^3$. These chambers provided 
an overall efficiency for charged track reconstruction 
of better than 95\% and a momentum 
resolution of approximately 3.5\% in the momentum range of interest
(less than 10~$\mbox{GeV}/\mbox{c}$).
Reconstructed tracks were used to determine the event topology 
(the assignment of tracks to vertices),
to reconstruct the vertex position and 
the track parameters at each vertex and, finally, to
identify the vertex type (primary, secondary, $\vo$, etc.).
A transition radiation detector~\cite{Bassompierre1,Bassompierre2} 
placed at the end of the active target was used for 
particle
identification.
A lead-glass electromagnetic 
calorimeter~\cite{Autiero:1996sp,Autiero:1998ya} located
downstream of the tracking region provided 
an energy resolution of $3.2\%/\sqrt{E \mbox{[GeV]} } \oplus 1\%$
for electromagnetic showers and was crucial to measure 
the total energy flow in neutrino interactions.
In addition, an iron absorber and a set of muon chambers located after 
the electromagnetic calorimeter were used for muon
identification, providing a muon detection 
efficiency of 97\% for momenta greater than 5~GeV/c.

Neutral strange particles were reconstructed and identified with high  
efficiency and purity using the $\vo$-like signature of their
decays~\cite{Astier:2000ax,Astier}.
Proton identification needed further development for the search
presented here using information from
the drift chambers, transition radiation detector and 
electromagnetic calorimeter.


The NOMAD Monte Carlo simulation (MC) is based on LEPTO 6.1
~\cite{Ingelman:1992ef} and JETSET 7.4~\cite{Sjostrand} generators
for neutrino interactions, and on a GEANT~\cite{GEANT} based program
for the detector response. 
The relevant JETSET parameters have been tuned in order to reproduce
the yields of strange particles measured in $\nu_\mu$ CC
interactions in NOMAD~\cite{Astier}. 
To define the parton content of the nucleon for the cross-section calculation
we have used the parton density distributions parametrized
in ~\cite{Alekhin}.

\section{\label{sec:selection} Event Selection}
We have analysed neutrino--nucleon
interactions of both charged (CC) and neutral current (NC) 
types. These events are selected with the requirements:
\begin{itemize}
\item[--] The reconstructed primary vertex should be within a fiducial
volume (FV) defined by $|x,y|<120$ cm,
$5<z<395$ cm (see Fig.~\ref{fig:nomad_detector} 
for the definition of the NOMAD coordinate system)
\item[--] There should be at least two charged tracks originating from the primary vertex;
\item[--] The visible hadronic energy should be larger than 3 GeV.
\end{itemize}
The $\nu_\mu$ CC events are identified requiring in addition:
\begin{itemize}
\item[--] The presence of an identified muon from the primary vertex.
\end{itemize}
The NC sample contains a contamination of about 30\% from unidentified
CC events. However, we do not apply further rejection against this background
in order not to reduce the statistics. The event purity
for the $\nu_\mu$ CC selection is $99.6\%$. The total sample amounts to 
about $1.5$ million neutrino events (see Table~\ref{tab:neutrino_events}).
\begin{table}[htb]
\begin{center}
     \begin{tabular}{||c|c|c|c||}
\hline \hline
               &  CC     & NC     &  CC+NC \\
\hline
     $N_{obs}$ &  785232 & 393539 &  1178771\\
\hline
     $N_{corr}$&  1017664& 481269 &  1498933\\
\hline \hline
    \end{tabular}
\caption{\label{tab:neutrino_events} Statistics of observed ($N_{obs}$) and efficiency corrected ($N_{corr}$) neutrino CC and NC events in the data.}
\end{center}
\end{table}

\subsection{\label{sec:k0_id} $\ko$ Identification}
\begin{figure*}[htb]
\begin{tabular}{cc}
\mbox{\epsfig{file=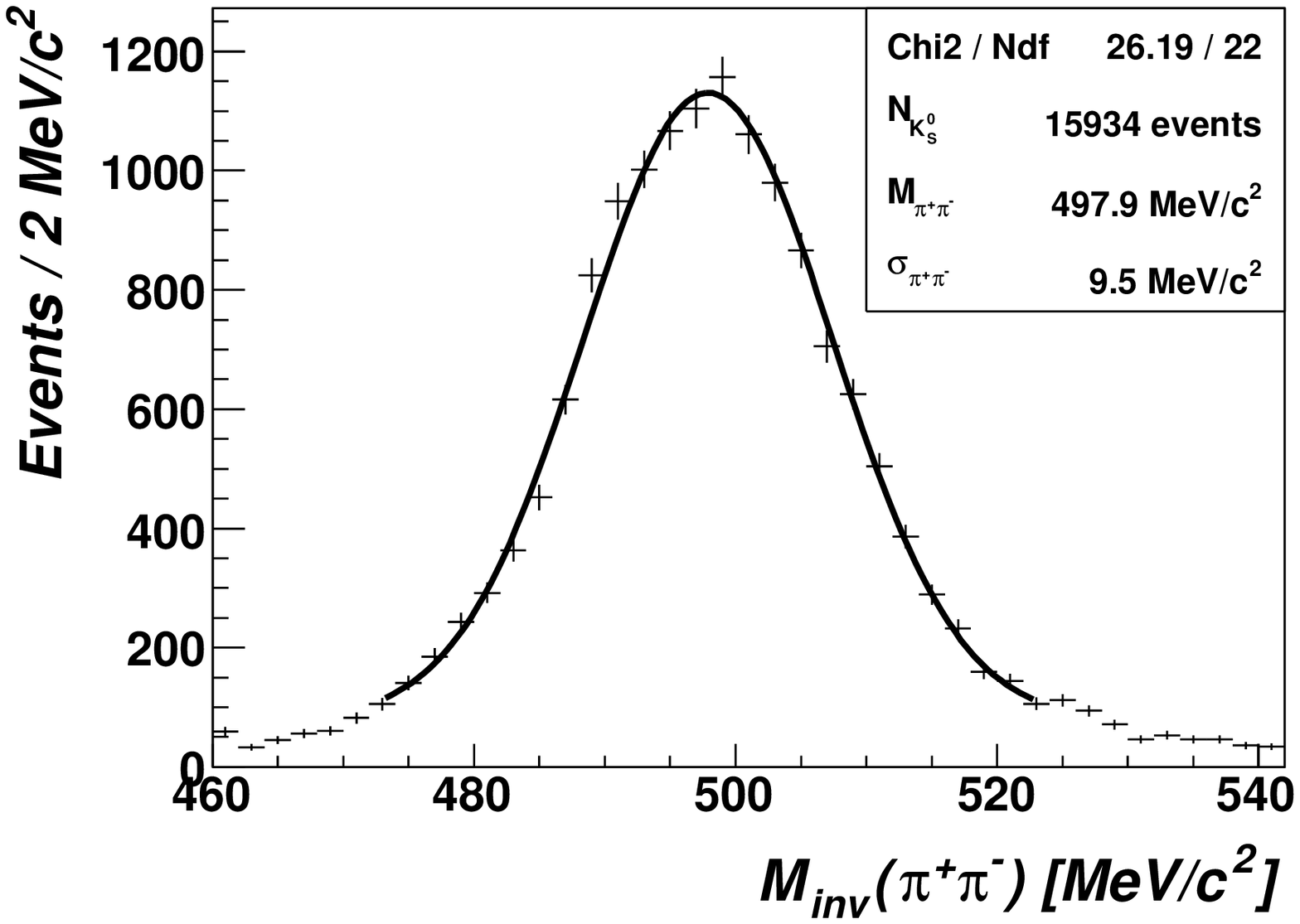,width=0.5\textwidth}} &
\mbox{\epsfig{file=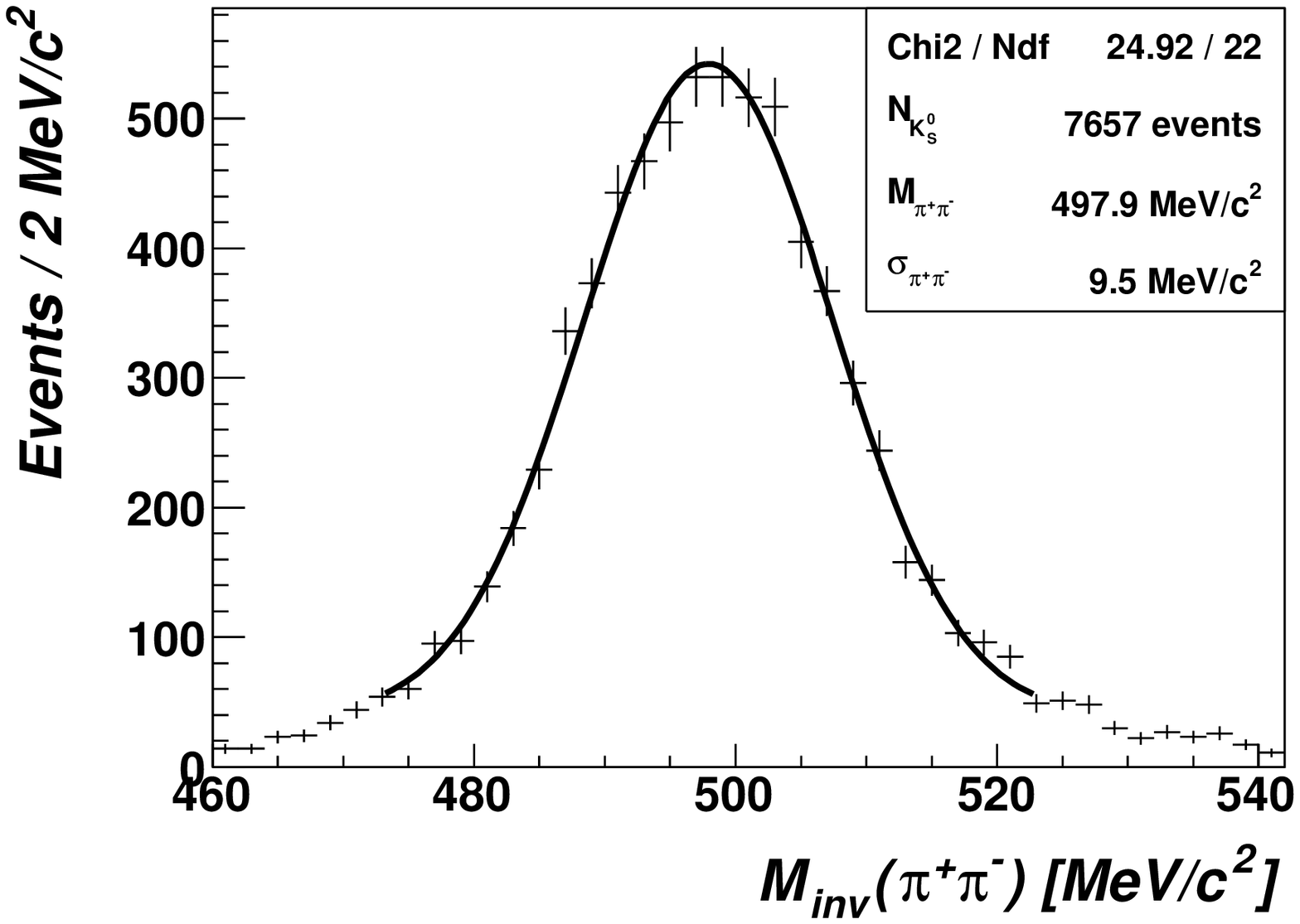,width=0.5\textwidth}}
\end{tabular}
\caption{\label{fig:k0mass} Reconstructed $\kodecay$ invariant mass
distribution in the $\nu_\mu$ CC (left) and $\nu_\mu$ NC (right) data
subsamples.}
\end{figure*}

$\ko$ mesons are identified through their $\vo$-like decay \\
$\kodecay$ using
a kinematic  constrained fit~\cite{Astier:2000ax,Astier}. 
With a purity of 97\% we identify
15934 and 7657 $\ko$ mesons in the CC and NC samples respectively,
thus yielding a total statistics of more than 23k $\ko$'s. The 
reconstructed $\kodecay$ invariant mass distribution in the $\nu_\mu$
CC (left) and $\nu_\mu$ NC (right)
subsamples are shown in  Fig.~\ref{fig:k0mass}. The two distributions
have the
same $\ko$ mass mean value, 497.9 MeV$/c^2$, in agreement with the PDG
value, and a width compatible with the expected experimental
resolution of $\sim$9.5 MeV$/c^2$.

\subsection{\label{sec:proton_id} Proton Reconstruction}
The identification of protons is the most difficult part of the present
analysis. As the NOMAD experiment does not include a dedicated
detector for proton identification, we developed a special procedure 
for this purpose.
The main background in the proton selection is the $\pi^+$ contamination 
since pions are about 2.5 times more abundant than protons. However
the $\pi^+$ contamination can be suppressed exploiting the differences in 
the behaviour of protons and pions
propagating through the NOMAD detector. We use three sub-detectors
which can provide substantial rejection factors against pions:
\begin{enumerate}
\item {\em The Drift Chambers (DC)}. A low energy proton 
ranges out faster than a pion of the same momentum.
Thus a correlation between the particle momentum and its path length
can be used as a discriminator between
protons and pions. The momentum interval of applicability of this method
is below 600 MeV$/c$.
\item {\em The Transition Radiation Detector (TRD)}. The energy deposition
of protons and pions  in the TRD is 
very different due to the larger proton ionization loss
for momenta below 1 GeV$/c$, allowing a good pion--proton separation
in this momentum interval. A modest discrimination is also possible
for momenta above 3 GeV$/c$ because of relativistic rise effects.
\item {\em The Electromagnetic Calorimeter (ECAL)}. The proton sample
can be cleaned further by taking into account
the different Cherenkov light emission of protons and pions of the same
momentum. 
\end{enumerate}

As a preliminary quality selection for the candidate proton track we require :
\begin{itemize}
\item[--] more than 7 hits on the DC track in order to have a 
reliable  fit;
\item[--] the distance between the primary vertex and the first hit of the 
track  to be smaller than 15 cm;
\item[--] the relative error on the track momentum to be smaller than 
0.3;
\item[--] the coordinate of the last hit of the stopped particle to be
within a fiducial volume (FV) defined as $|x|<120$ cm, $-110<y<100$ cm,
$35<z<380$ cm, which is smaller than the FV used in the event selection
in order to reduce edge effects;
\item[--] the particles reaching the TRD to cross at least 6 TRD
	planes (out of 9 in total) in order to allow for the TRD 
identification algorithm.
\end{itemize}

The initial sample of positively charged tracks is split into two subsamples:
those that passed the quality criteria and those that did not.
The identification cuts can be applied only to the positive tracks
that passed the quality criteria.
Charged particles were all assumed to be pions in the standard
reconstruction program, resulting in a
systematic underestimation of the reconstructed proton momentum ($p_{rec}$)
at small momenta. The difference $\Delta p$ between the true and
reconstructed proton momentum was studied with the help of the MC, and 
the following parametrization was found to describe the effect :
\begin{equation}
 \label{eq:correction}
\Delta p = 0.33 \cdot e^{- \frac{3.5 p_{rec}}{GeV/c}}\mbox{ GeV/c}.
\end{equation}

\begin{figure}[htb]
 \begin{center}
 \epsfig{file=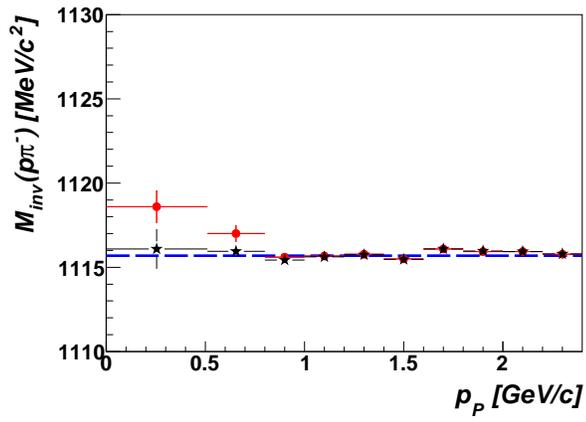,width=\linewidth}
 \end{center}
\caption {The mean value of the invariant mass of  $p \pi^-$ pairs from
identified $\Lambda$'s as a function of the reconstructed proton momentum
with no momentum correction (full circles) and with momentum correction
(stars).}
\label{fig:invmass_lambda}
\end{figure}

The difference $\Delta p$ decreases with $p_{rec}$, becoming negligible
at about 0.8 GeV/c. 
The reconstructed proton momentum was corrected accordingly.
The effect of this proton correction was tested on a reconstructed sample of
$\lam$ hyperons identified by their displaced decay vertex.
Fig.~\ref{fig:invmass_lambda} displays the mean value of the invariant
$p \pi^-$ mass as a function of the reconstructed proton momentum without
and with the $\Delta p$ correction.
There is an improvement in the reconstructed $\lam$ mass when
correcting the reconstructed proton momentum, especially at low momenta.

\section{\label{sec:background} The background} 
Random $\ko$--proton pairs produce combinatorial background in the
    $\ko$p invariant mass ($M(\ko p)\equiv M$) distribution. Understanding the shape of this background
    is crucial in the search for a possible $\Theta^+$ signal. We studied this background in three
    different ways:
\begin{enumerate}
\item MC events contain no  $\Theta^+$ and could be used, therefore,to study
the background for this analysis. However, the small fraction of
proton--$\ko$ pairs with an invariant mass in the interesting mass region 
would require a very large sample of MC events to reduce statistical
fluctuations.
\item We combined  protons and $\ko's$ from different events in the data,
thus making {\em fake pairs}, paying 
special attention that the original data distributions of multiplicity, proton
and kaon momenta, and their relative opening angle, were well reproduced 
in the final {\em fake pair} sample.
\item A polynomial fit to the $M$ distribution of the data themselves,
excluding the $\Theta^+$ mass region, can also be used to describe the
background for the $\Theta^+$ search.
\end{enumerate}

\begin{figure*}[htb]
\begin{center}
\begin{tabular}{cc}
\epsfig{file=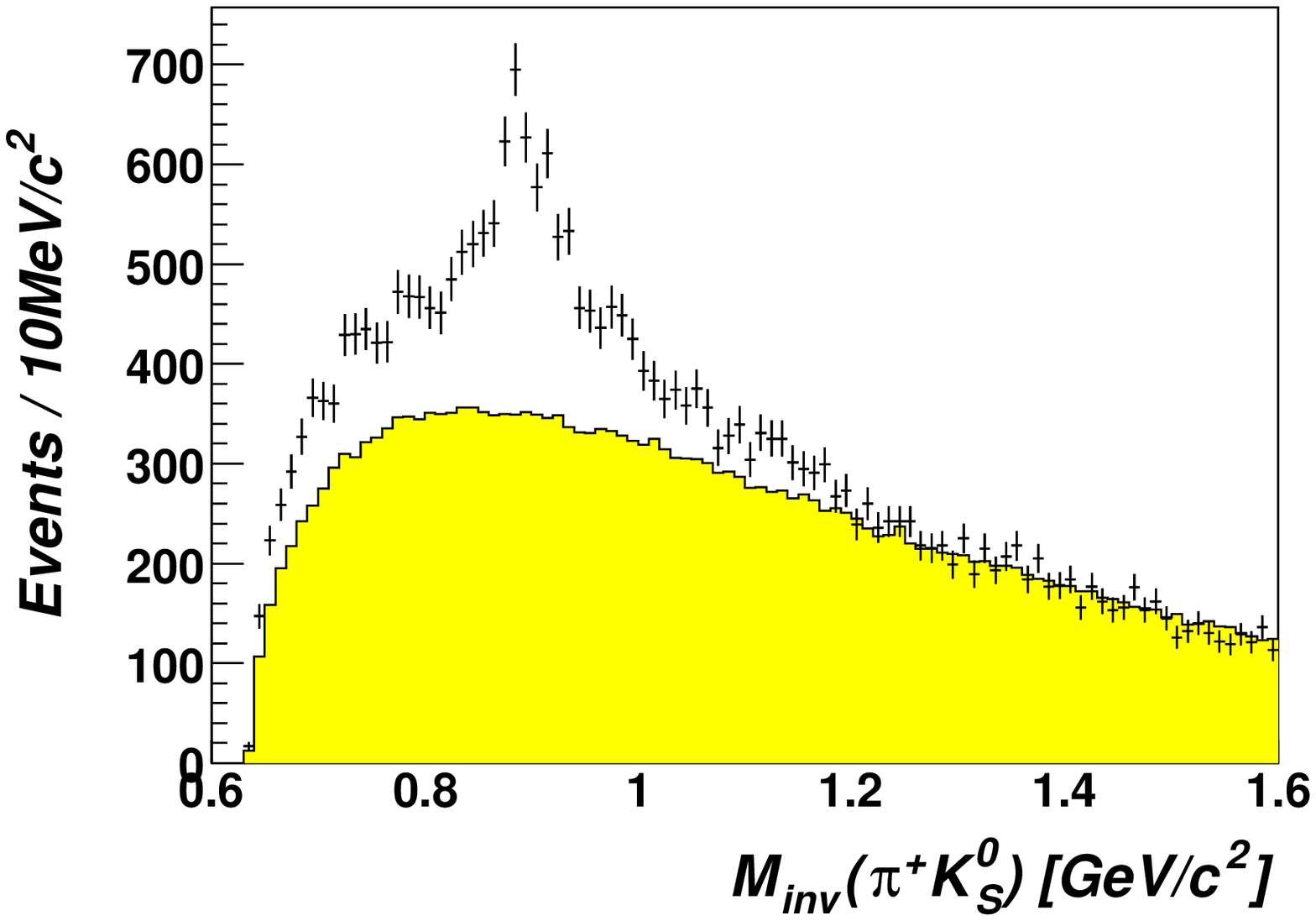,width=0.45\textwidth}&
\epsfig{file=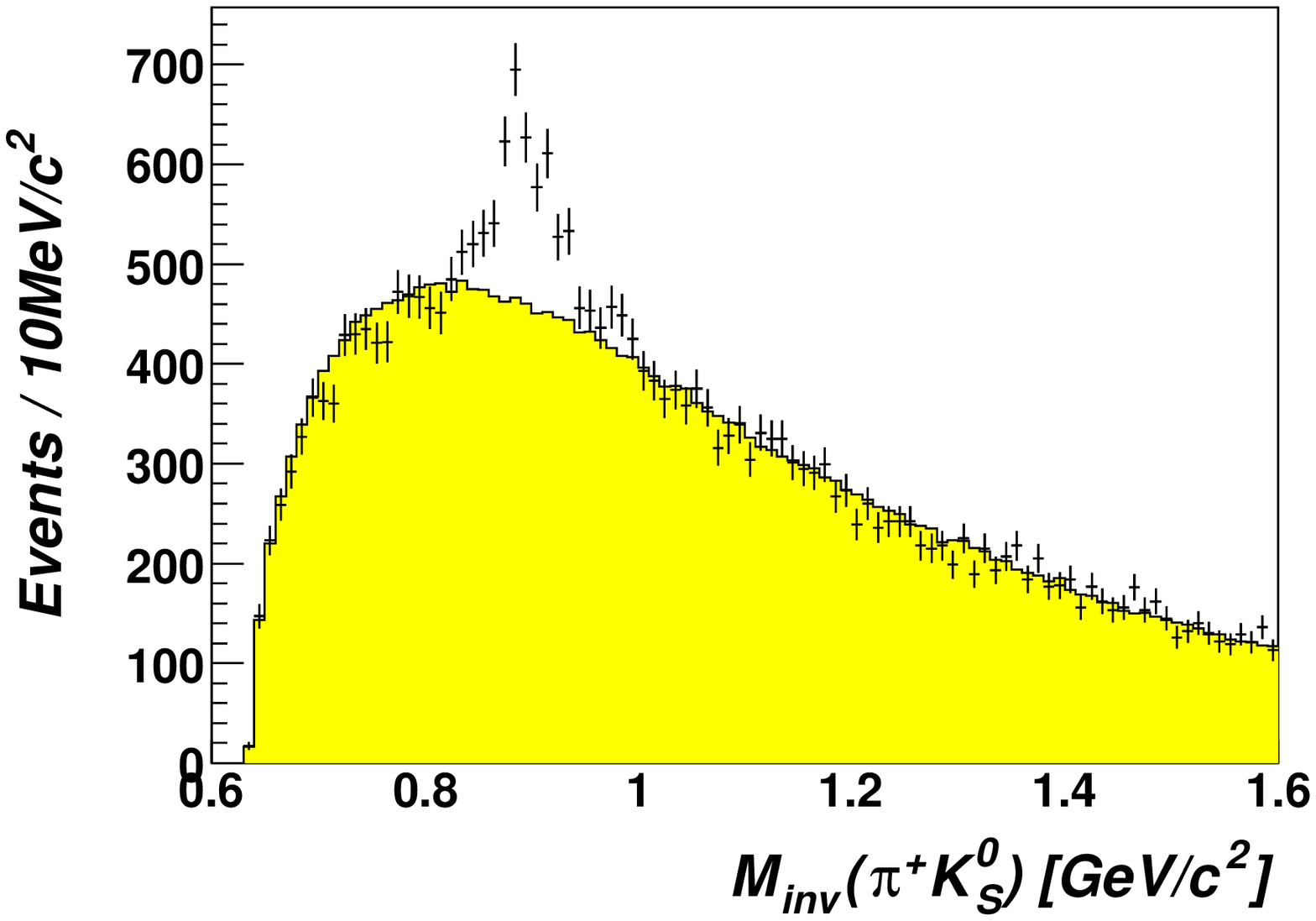,width=0.45\textwidth}\\
\end{tabular}
\caption{\label{fig:kstar_back} $\ko\pi^+$ invariant mass distribution in the
data, superimposed with the background generated without hadronic jet rotation
 (left), and with  hadronic jet rotation (right).}
\end{center}
\end{figure*}

The {\em fake pair} technique is extensively used in the literature.
However, it is necessary to ensure that the two independent events 
used in the mixing have similar hadronic jet momenta both in magnitude
and direction. If two events with different jet momenta are mixed, 
then the {\em fake pair} technique systematically underestimates the
background at small $M(\ko p)$ invariant masses. This is illustrated
in the left panel of Fig.~\ref{fig:kstar_back} which shows the
$\ko\pi^+$ invariant mass distribution in the data, superimposed with
the background generated without changing the hadronic jet directions.
The background is normalized to the data at $M({\ko\pi^+})=1400$ MeV$/c^2$.
The data show a clear $K^{*+}$ peak at $\sim 890$ MeV$/c^2$. However,
the background under this peak is obviously underestimated. Therefore,
in our procedure we first rotate each data event such that the
hadronic jet momentum is aligned along the z-axis. We then select events
respecting the original multiplicity of positive tracks and $\ko$ in
the data and make random pairs. The right panel of Fig.~\ref{fig:kstar_back}
shows the same $\ko\pi^+$ invariant mass distribution in the data
superimposed with the background generated according to our procedure.
There is now a clear agreement with the data distribution, except for
the $K^{*+}$ peak which the {\em fake pair} technique cannot reproduce.

\begin{figure*}[htb]
\begin{center}
\begin{tabular}{cc}
\epsfig{file=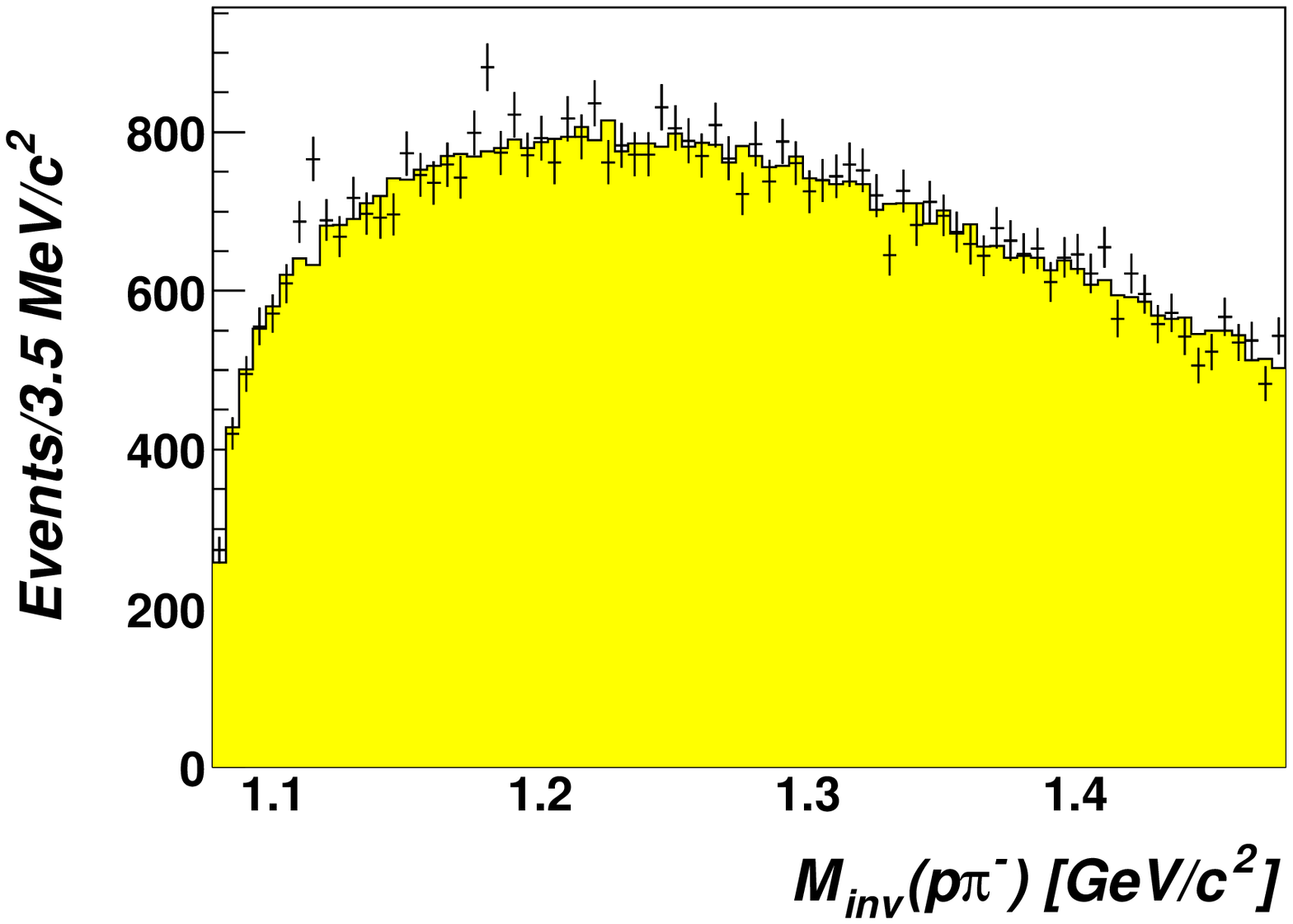,width=0.45\textwidth}&
\epsfig{file=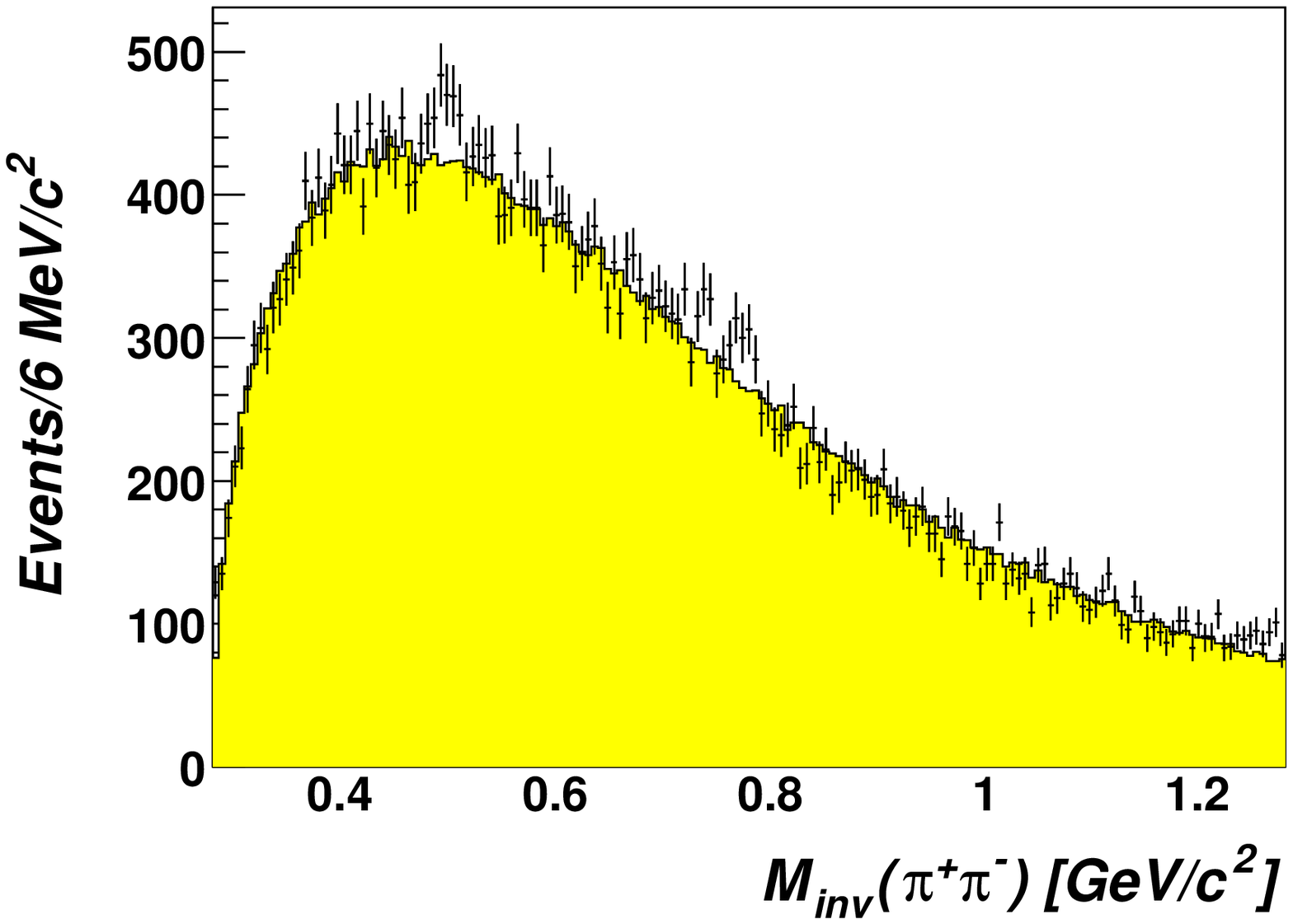,width=0.45\textwidth}\\
\end{tabular}
\caption{\label{fig:k0_lam_back} $p\pi^-$ (left) and $\pi^+\pi^-$ (right)
invariant mass distribution (points with error bars) superimposed to the
estimated background.}
\end{center}
\end{figure*}

We have checked the background generation procedure on samples of 
$\lam\to p\pi^-$ and $\ko\to \pi^+\pi^-$ events, using only
tracks originating from the primary interaction vertex in order to artificially increase the
background. Fig.~\ref{fig:k0_lam_back} shows
$p\pi^-$ and $\pi^+\pi^-$ invariant mass distributions superimposed
with the predicted background estimated using our {\em fake pair} procedure.
From these plots we conclude that our background generation procedure
provides a realistic background estimate in these cases as well.

It is worth mentioning however that the {\em fake pair} tecnique does not take into account contributions from resonances which introduce {\em correlations} in $M$ different from those generated by making random pairs. $\Sigma(1660)-\Sigma(2250)$ decaying into proton and kaon and $\Kdecay$ decays with the pion taken as the proton might be important sources of distortion of the background shape. With help of MC we find a negligible contribution of $\Sigma(1660)-\Sigma(2250)$ resonances to $\ko p$  invariant mass distribution while $\Kdecay$ decays increase the background by 5-10\% at small $M\sim 1500$ MeV$/c^2$. Therefore we renormalize the background distribution obtained by {\em fake pair} procedure by a ratio of two $\ko p$ distributions obtained with help of MC with and without $\Kdecay$ decays. Finally the obtained background distribution is normalized to the data distribution at $M>1650$ MeV$/c^2$. Fig.~\ref{fig:data_vs_background_blind} shows the invariant mass distributions of combinations of a positively charged track, assumed to be a proton, and a $\ko$ for the data and for the fake pair background, without using
proton identification and with ``optimal`` proton identification. The ``signal`` interval $1510< M < 1550$ MeV$/c^2$ is {\em excluded} in the data. There is good agreement between the shapes of the data and background distributions. Polynomial fits of the data excluding the ``signal`` interval $1510< M < 1550$ MeV$/c^2$ is also shown as dashed curves. There is a reasonable agreement of the background shapes obtained by {\em fake pair} procedure and by a polynomial fit of the data.

\section{\label{sec:theta_tools}$\Theta^+$ analysis tools}
\subsection{\label{sec:proton-strategy} The proton identification strategy} 
The $\Theta^+$ signal is expected to appear as a narrow peak in the invariant
mass distribution of $\ko$--proton pairs. $\ko$ are identified using their
$\vo$-like signature (see Sec.~\ref{sec:k0_id}). To separate protons
from $\pi^+$, for each positively charged track we build likelihoods
under the proton and $\pi^+$ hypothesis using the information from DC,
TRD, and ECAL (see Sec. 3.2), and we take their ratios 
${\cal L}_{DC}$, ${\cal L}_{TRD}$, ${\cal L}_{ECAL}$:

\begin{equation}
\label{eq:likelihoods}
\begin{aligned}
&{\cal L}_{DC}(p,L),                 && L - \mbox { track length} & \\
&{\cal L}_{TRD}(p,\epsilon_{TRD}),   && \epsilon_{TRD} - \mbox{ energy release in TRD} \\
&{\cal L}_{ECAL}(p,\epsilon_{ECAL}), && \epsilon_{ECAL} -\mbox{ energy release in ECAL}\\
&\phantom{X}                         && p - \mbox{ track momentum. }                   \\
\end{aligned}
\end{equation}

We optimize the cuts for the proton identification likelihood ratios
maximizing the sensitivity to the expected $\Theta^+$ signal. These 
``optimal`` cuts are not necessarily those which maximize the purity of
the proton sample.

The best approach for tuning the proton identification cuts would be to
maximize the sensitivity using a detailed Monte Carlo for $\Theta^+$
production. However, given the poor knowledge on the properties of this
particle, there is no available MC generator describing the production
of exotic baryons. We create, therefore, ``fake`` $\Theta^+$ states 
in the NOMAD event generator by using pairs of protons and $\ko$ with
invariant mass close to the mass of $\Theta^+$ state. However, in this
approach the momentum distribution of these ``fake`` $\Theta^+$ states
is determined by the momentum distribution of protons and $\ko$ from the
primary vertex. This can result in wrong ``optimal`` cuts if the true
momentum distribution of $\Theta^+$ particles is very different.  We try to
avoid this problem by subdividing the original MC sample into several
narrow bins of $x_F$ and optimizing the cuts for {\em each} $x_F$ interval
independently. The $x_F$ variable is defined as the ratio of the longitudinal
projection of the $\Theta^+$ momentum on the hadronic jet momentum to the
hadronic jet energy in the hadronic center-of-mass frame. The variable
$x_F$ is in the range $(-1,1)$ with negative (positive) values often called
the {\em target (current) } fragmentation regions. 

The procedure of tuning the proton identification cuts is then as follows:
\begin{itemize}
\item We build ``fake`` $\Theta^+$ states by taking $\ko$--proton pairs
with $1510 < M < 1550$ MeV$/c^2$. Assuming no $\Theta^+$ polarization,
a flat distribution of $\cos\theta^*$, where $\theta^*$ is the angle between
the proton momentum in the $\Theta^+$ rest frame and the $\Theta^+$ momentum
in the laboratory. We reweight the $\cos\theta^*$ distribution so obtained
to make it flat. This is our MC ``signal``.
\item Any other combination of a $\ko$ and a positive track not identified
as a proton, but with an assigned proton mass, is taken as the MC background
if its invariant mass M falls in the same mass interval.
\item We split the ``fake`` $\Theta^+$ states into several intervals of
positive track momentum. We vary the cuts on
${\cal L}_{DC}$, ${\cal L}_{TRD}$, ${\cal L}_{ECAL}$
simultaneously in each interval and find those cuts which maximize the
$signal/\sqrt{background}$ ratio.
\end{itemize}

\begin{figure*}[htb]
 \begin{center}
    \begin{tabular}{cc}
      \epsfig{file=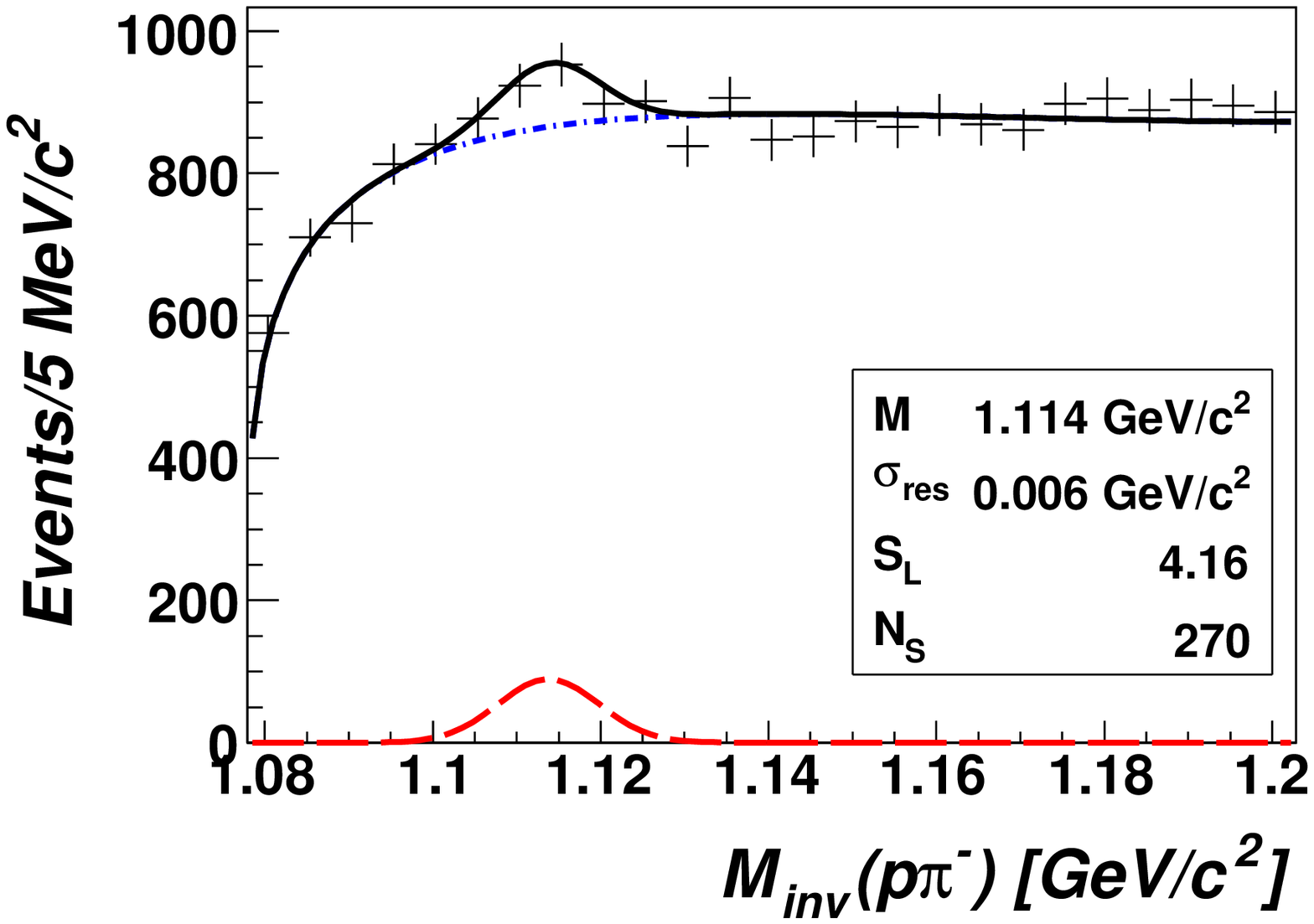,width=0.4\linewidth}&
      \epsfig{file=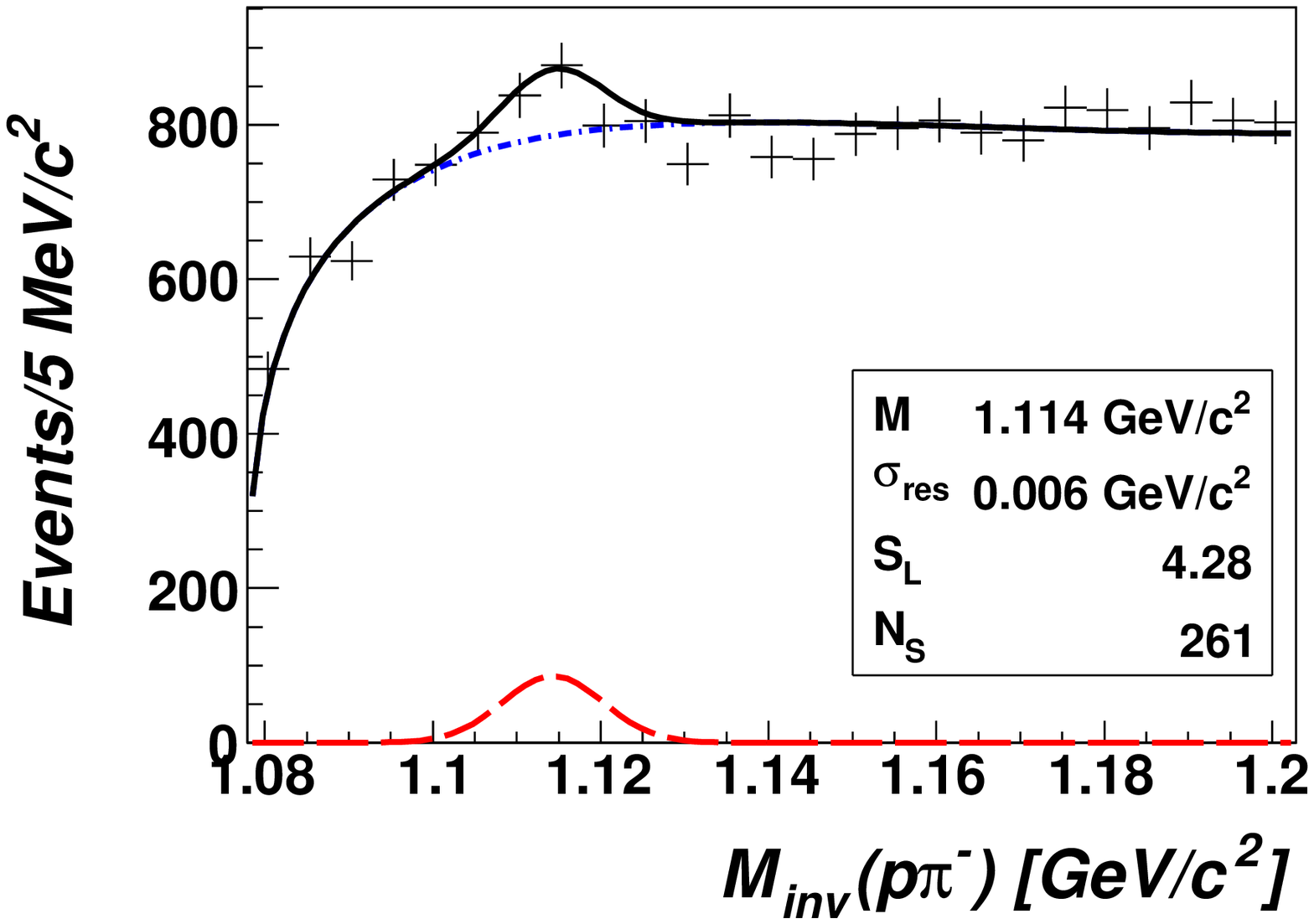,width=0.4\linewidth}\\
      \epsfig{file=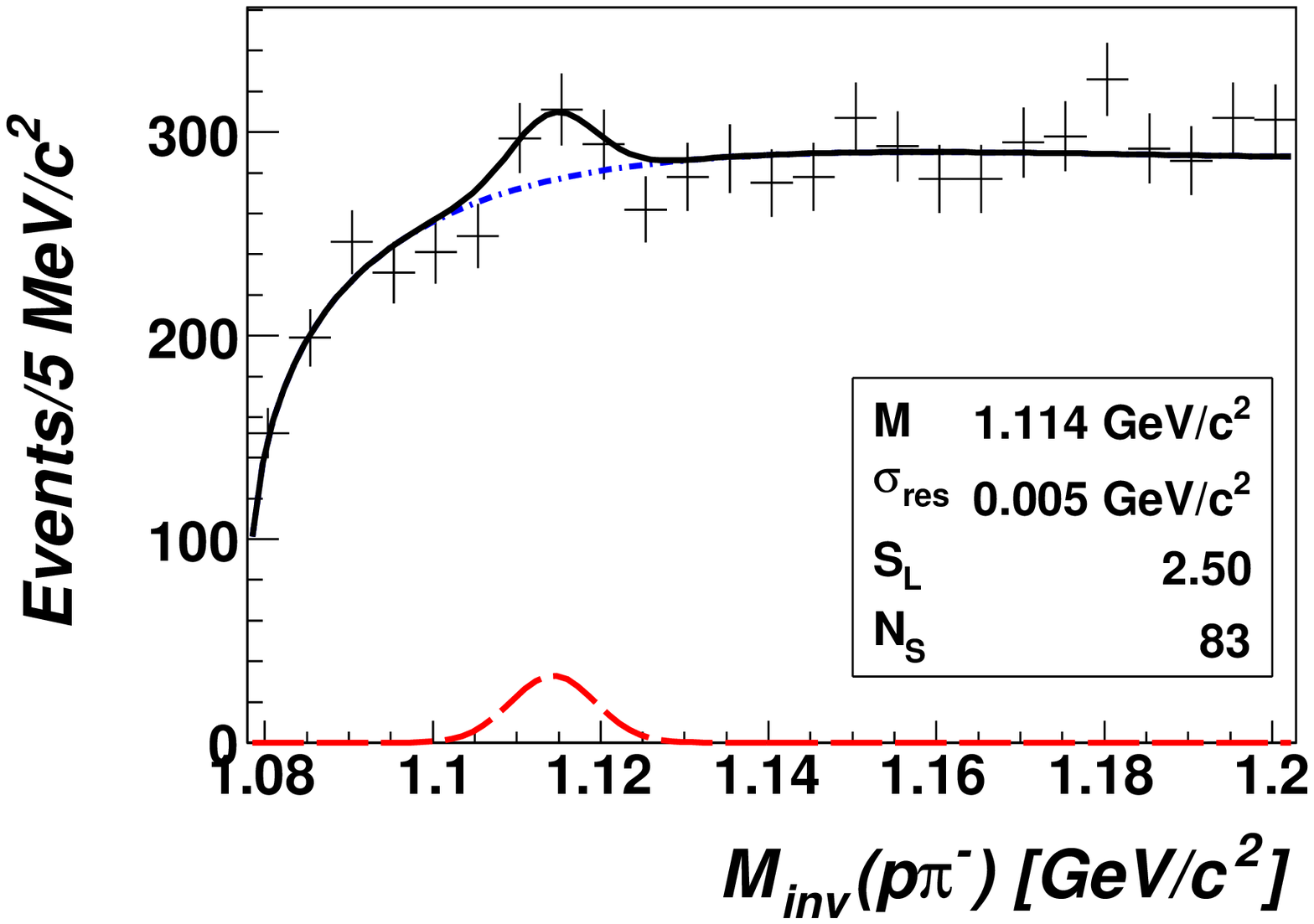,width=0.4\linewidth}&
      \epsfig{file=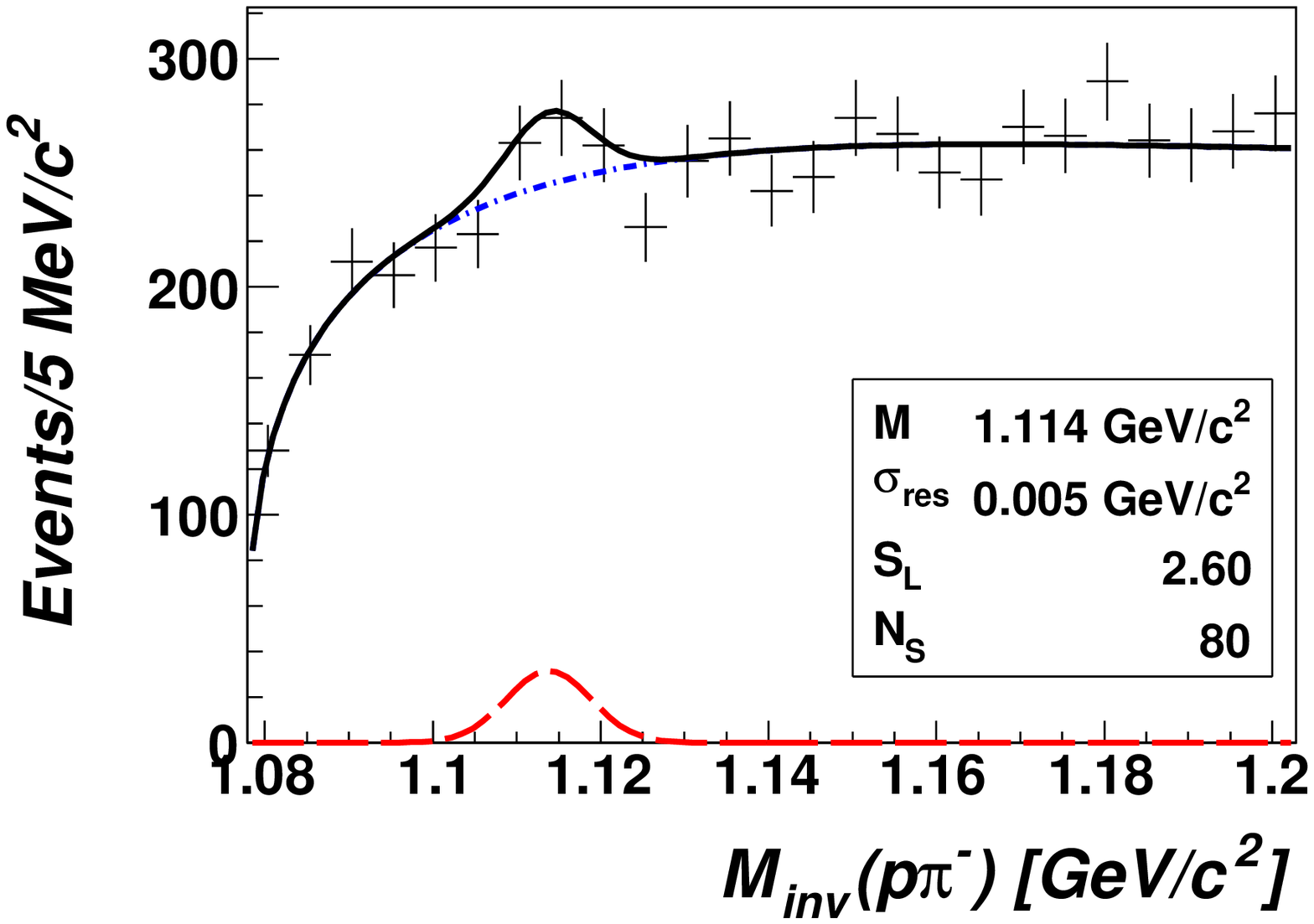,width=0.4\linewidth}\\
    \end{tabular}
 \end{center}
\caption {Invariant mass distributions of proton--$\pi^-$ pairs for $-0.6 < x_F < -0.3$ interval.
(Left, up) MC, no proton identification; (Right, up) MC, with ``optimal``
proton identification; (Left, down) data, no proton identification;
(Right, down) data, with ``optimal`` proton identification. The curves
represent the predicted background and the amount of $\lamdecay$ signal
maximizing the $signal/\sqrt{background}$ ratio (see Sec. 5.1).}
\label{fig:lambda-xF-optimum}
\end{figure*}

We check this procedure on a sample of $\lamdecay$ events. 
Fig.~\ref{fig:lambda-xF-optimum} displays the invariant mass distributions
of proton--$\pi^-$ pairs in both MC and data without proton identification
and with ``optimal`` for the $\lamdecay$ observation proton identification, for $-0.6 < x_F < -0.3$.
With ``optimal`` proton identification the significance of the
$\lamdecay$ signal increases in both MC and data samples.

\begin{table}[htb]
\begin{center}
\begin{tabular}{||c|c|c|c|c||}
\hline\hline
                      & \multicolumn{2}{|c|}{$N({p\ko})$}    & \multicolumn{2}{|c||}{purity (in \%)}\\
\cline{2-5}
                      & all  &  ``signal`` & all &  ``signal`` \\
\hline
no ID          & 53463   & 1856                     & 23      & 16.4\\
\hline
``optimal`` ID & 40561   & 1090                     & 27.8    & 22.1\\
\hline\hline
\end{tabular}
\end{center}
\caption{\label{tab:events_purity} Numbers of $p\ko$ pairs and purity of proton samples in the data for two subsets of events: without proton identification and with ``optimal`` proton identification. These numbers are shown for all entries and for ``signal`` region: $1510<M<1550$ MeV$/c^2$.}
\end{table}
In Tab.~\ref{tab:events_purity} we show numbers of $p\ko$ pairs and purity of proton samples in the data for two subsets of events: without proton identification and with ``optimal`` proton identification. These numbers are shown for all entries and for ``signal`` region ($1510<M<1550$ MeV$/c^2$).
\subsection{\label{sec:massresolution} The $p\ko$ mass resolution} 

The expected mass resolution of the $p\ko$ pair is estimated as follows.


\begin{itemize}
\item For MC events we calculate the invariant masses of the generated and
reconstructed $p K^0_S$ pairs, and we fit the distribution of the difference
between the two values by a Gaussian whose width is taken as the mass
resolution (method ``A``).
\item Using the measured momenta of the proton ($\vec{p}_1$) and of the
$\ko$  ($\vec{p}_2$), the angle $\theta$ between  $\vec{p}_1$ and $\vec{p}_2$,
and the associated errors $\sigma(\vec{p}_1)$ and $\sigma(\vec{p}_2)$ we find (neglecting errors in $\cos\theta$): 
\begin{equation}
\begin{aligned}
M^2_{inv} \ \sigma^2(M_{inv}) = & \left(\frac{E_2}{E_1} p_1 - p_2 \ cos \theta\right)^2 \sigma^2(p_1) + \\
                                & \left(\frac{E_1}{E_2} p_2 - p_1 \ cos \theta\right)^2 \sigma^2(p_2).
\end{aligned}
\end{equation}
This method,``B``, can be applied to both MC and data events.
\end{itemize}

\begin{figure}[htb]
 \begin{center}
 \epsfig{file=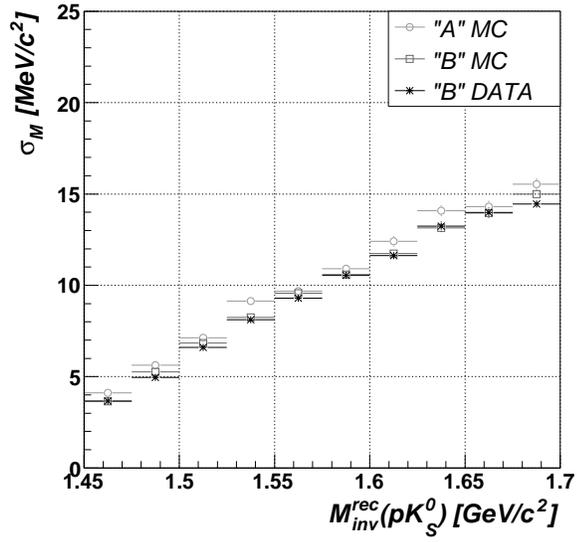,width=\linewidth}
 \end{center}
\caption {Expected invariant mass resolution of proton--$\ko$ pairs as a
function of the invariant mass for method ``A`` (MC only), and for
method ``B`` (MC and data). See text for details.}
\label{fig:resolutionmass2}
\end{figure}

Fig.~\ref{fig:resolutionmass2} displays the expected mass resolution
of $p\ko$ pairs as a function of their reconstructed invariant mass,
as obtained using method ``A`` (MC only), or method ``B`` (for both MC and data).
The results agree well with each other and predict a resolution of about
8.8 MeV$/c^2$ at the $\Theta^+$ mass (1530 MeV$/c^2$).

\subsection{\label{sec:stat_analysis} The statistical analysis} 
An estimation of the signal significance in the data is performed as follows:
\begin{enumerate}
\item A possible difference in the proton $\cos\theta^*$ distribution for
the signal and background is exploited to improve the signal
sensitivity. We take all $\ko$--proton pairs with $1510 < M < 1550$ MeV$/c^2$,
and we split them into 10 intervals with similar statistics: five mass
intervals with $\cos\theta^*$ in the interval $[-1, -0.5)$, and another
five mass intervals with $\cos\theta^*$ in the interval $[-0.5, 1]$.
The total mass interval ($1510 < M < 1550$ MeV$/c^2$) covers well the
expected $\Theta^+$ mass. The mass bin width, 10 MeV$/c^2$, is comparable
to the expected invariant mass resolution of $\ko$--proton pairs. 
\item We compute two likelihoods:
 \begin{equation}
\label{eq:likelihoods1}
\begin{aligned}
\ln L_B &=&& \sum_{i=1,10} \left(-b_i + n_i\cdot \ln b_i\right)\\
\ln L_{B+S} &=&& \sum_{i=1,10} \left(-b_i -s_i + n_i\cdot \ln \left(b_i + s_i\right) \right)\\
\end{aligned}
\end{equation}
where $b_i$, $s_i$, $n_i$ are the number of predicted background and signal
events, and observed data events in the $i$-th bin.
\item We compute the signal statistical significance as:
\begin{equation}
\label{eq:likelihoods2}
S_L = \sqrt{2 \left(\ln L_{B+S} - \ln L_B\right)}
\end{equation}
\item We find the resonance mass position $M$ and Breit-Wigner width
$\Gamma$ and the number of signal events $N_s$ which maximize $S_L$.
\end{enumerate}
For the background we use the procedure decribed in Sec.~\ref{sec:background}. 
The signal is modeled by a Breit-Wigner distorted by a Gaussian resolution
with $\sigma=8.8$ MeV$/c^2$. This algorithm was checked on several generated
distributions containing a Breit-Wigner signal of width $\Gamma$ distorted
by a Gaussian resolution of width $\sigma$ and superimposed on a fluctuating
background. We considered three cases, $\sigma\ll\Gamma$, $\sigma=\Gamma$,
$\sigma\gg\Gamma$, and found that in all cases the procedure of maximizing
$S_L$ correctly determined the number of signal events and $\Gamma$
(with $\Gamma$ around zero for the case $\sigma\gg\Gamma$).

\subsection{\label{sec:openbox} Opening the box}
\begin{figure*}[htb]
\begin{center}
\begin{tabular}{cc}
\epsfig{file=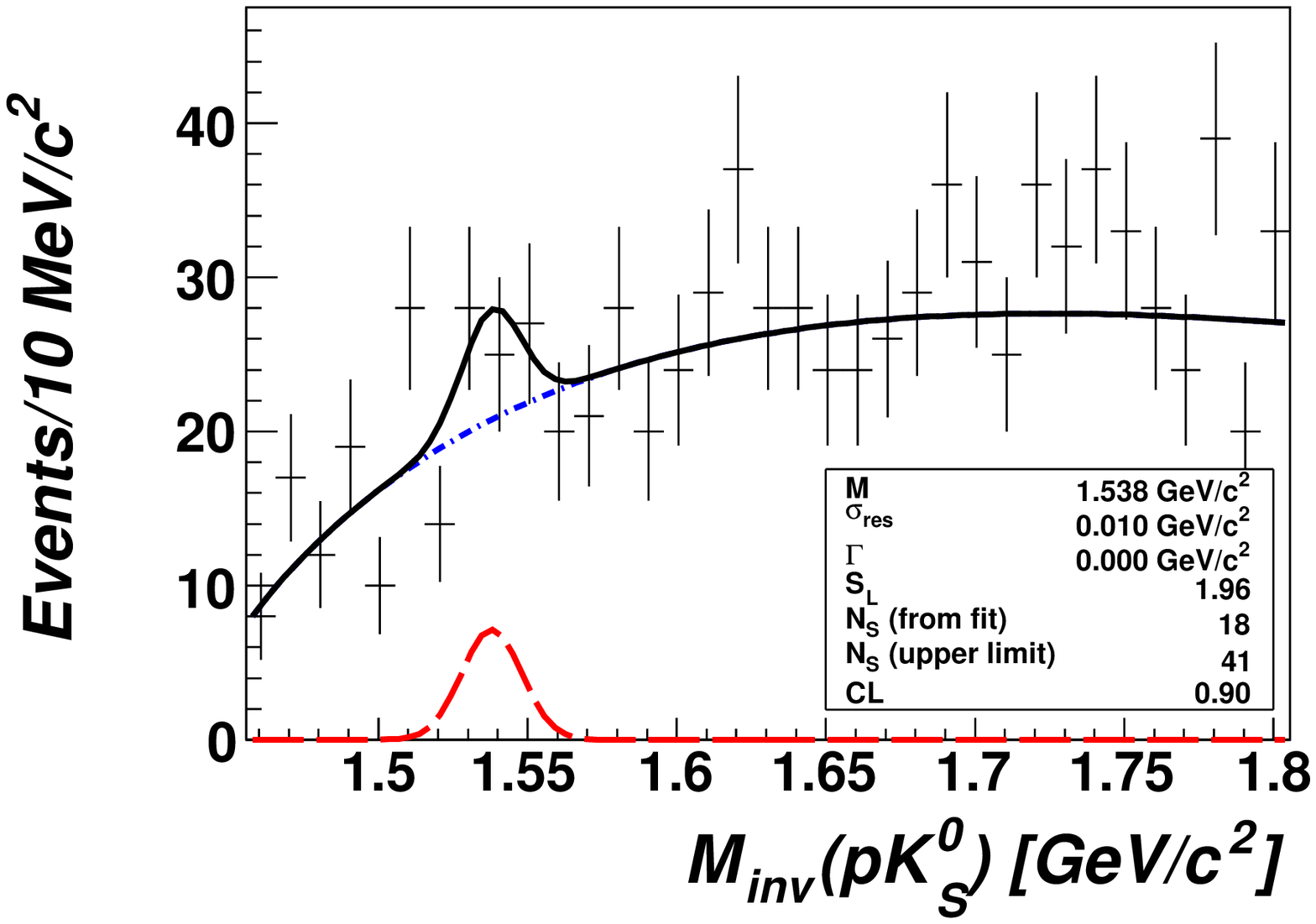,width=0.45\textwidth}&
\epsfig{file=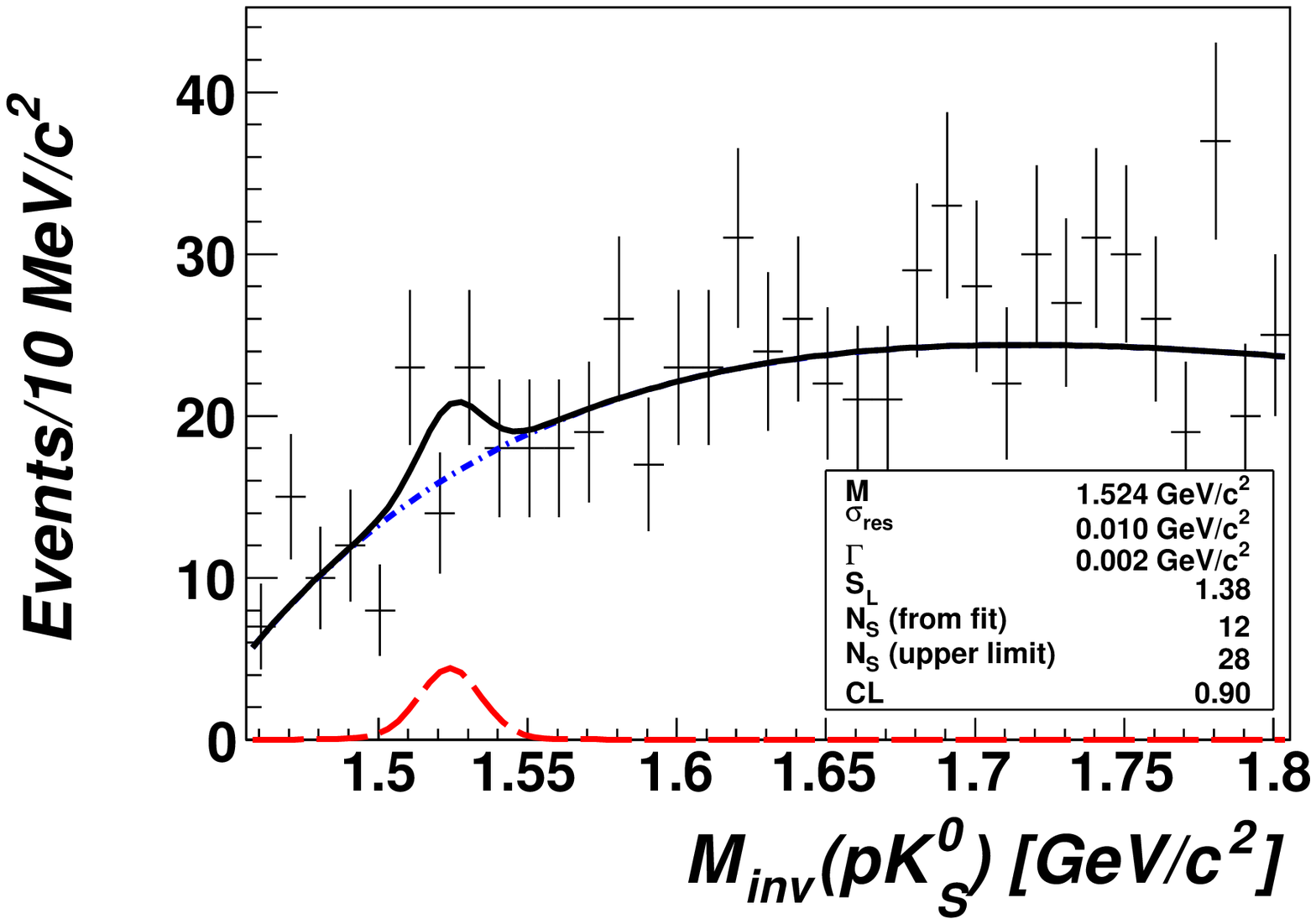,width=0.45\textwidth}\\
\end{tabular}
\caption{\label{fig:openbox-xf1} Invariant mass distributions of pairs of a
positively charged track (assumed to be a proton) and a $\ko$ in the data,
for $-1<x_F<-0.6$.
(Left) No proton identification; (Right) Optimized proton identification.
The curves represent the predicted background and the amount of $\Theta^+$
signal maximizing $S_L$ (see Sec. 5.3).}
\end{center}
\end{figure*}

\begin{figure*}[htb]
\begin{center}
\begin{tabular}{cc}
\epsfig{file=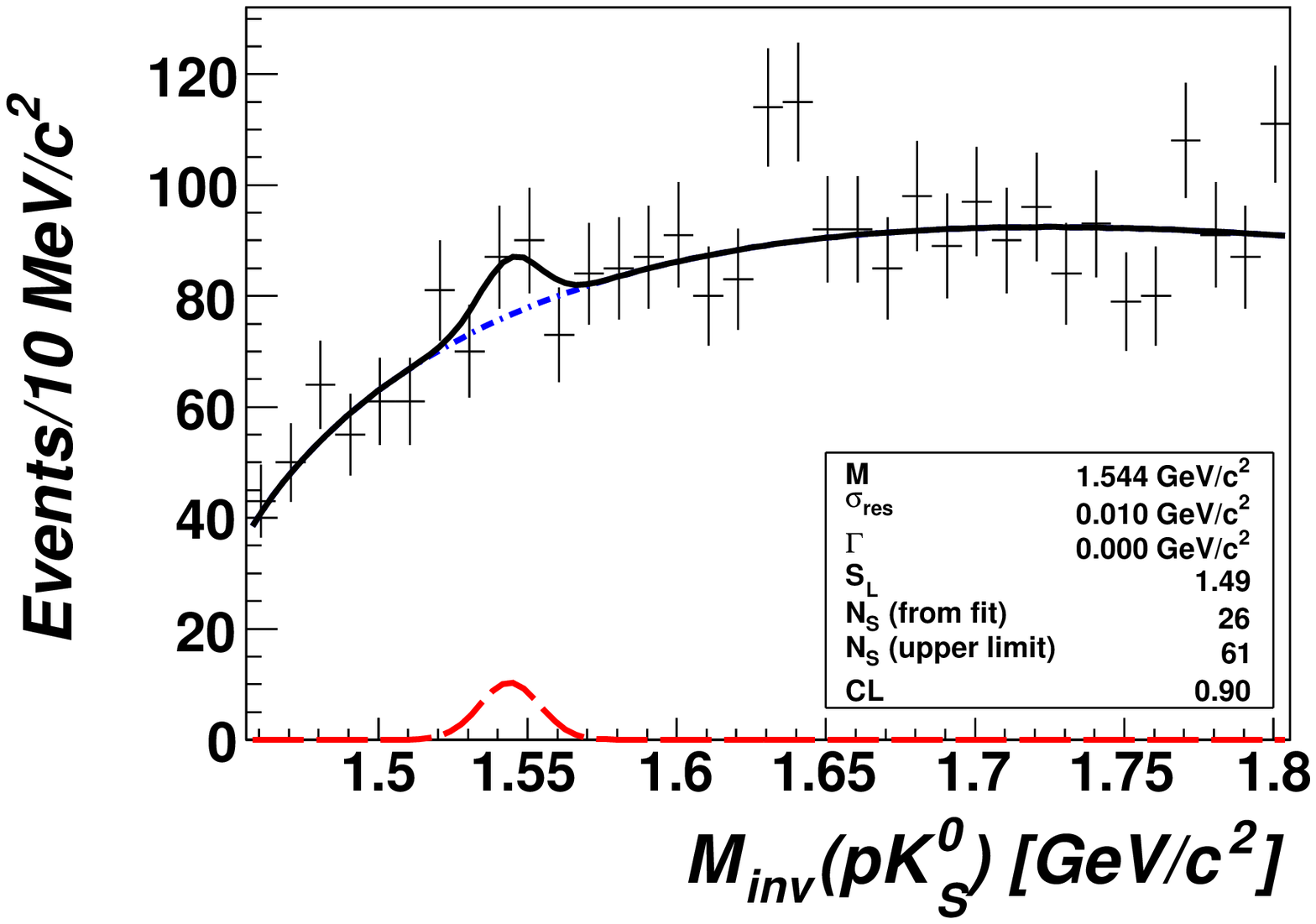,width=0.45\textwidth}&
\epsfig{file=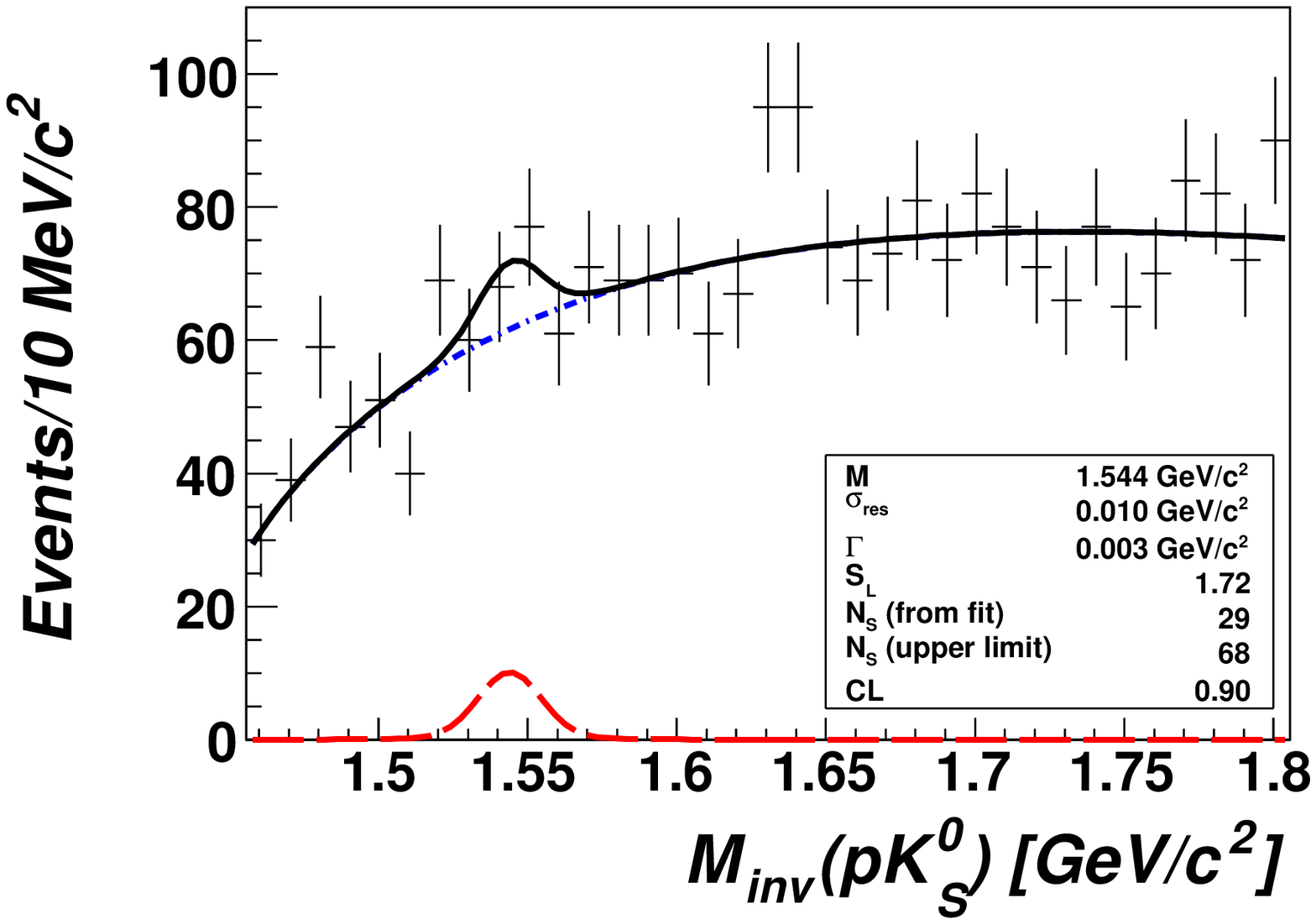,width=0.45\textwidth}\\
\end{tabular}
\caption{\label{fig:openbox-xf2} Invariant mass distributions of pairs of a
positively charged track (assumed to be a proton) and a $\ko$ in the data,
for $-0.6<x_F<-0.3$.
(Left) No proton identification; (Right) Optimized proton identification.
The curves represent the predicted background and the amount of $\Theta^+$
signal maximizing $S_L$ (see Sec. 5.3).}
\end{center}
\end{figure*}

\begin{figure*}[htb]
\begin{center}
\begin{tabular}{cc}
\epsfig{file=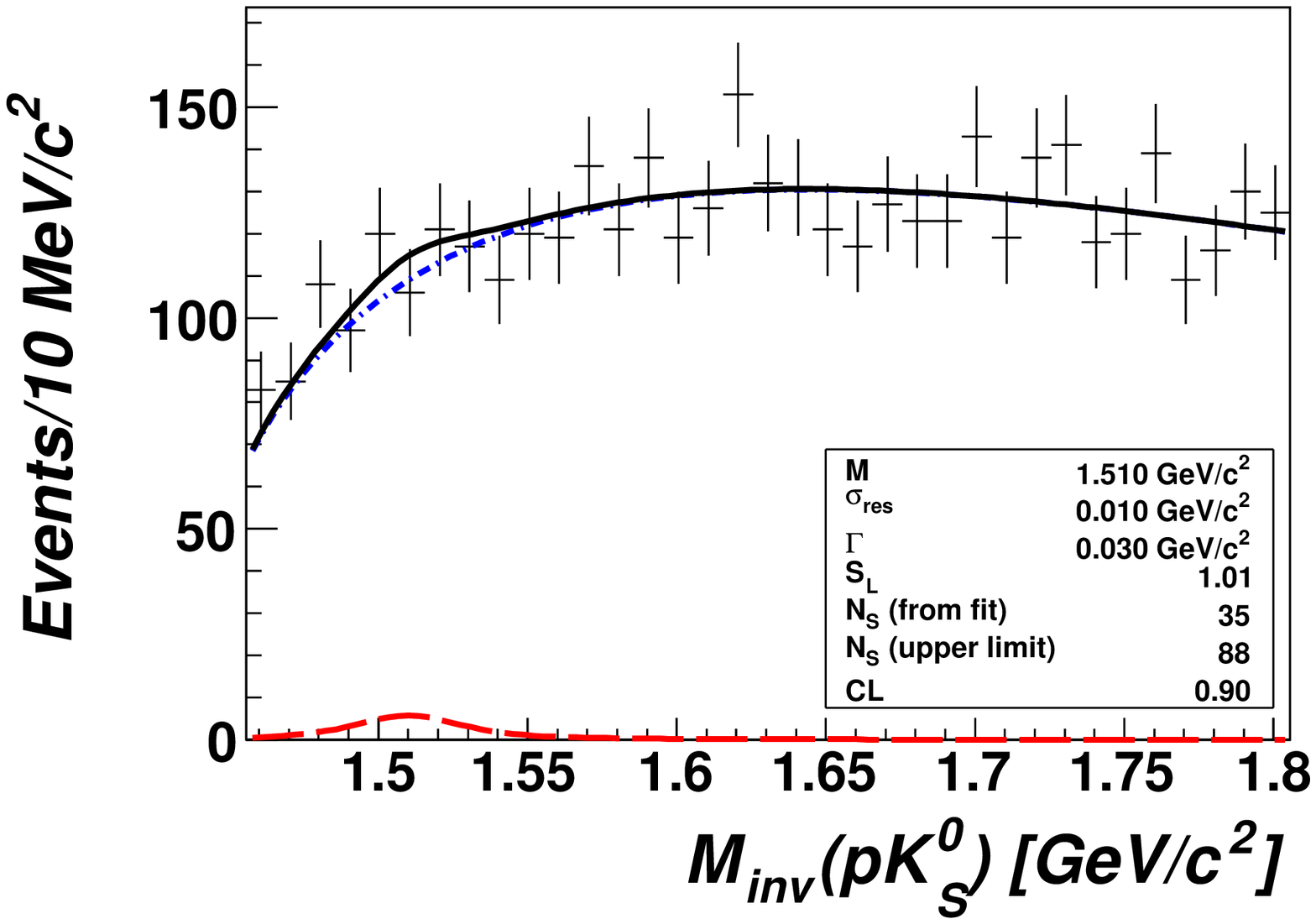,width=0.45\textwidth}&
\epsfig{file=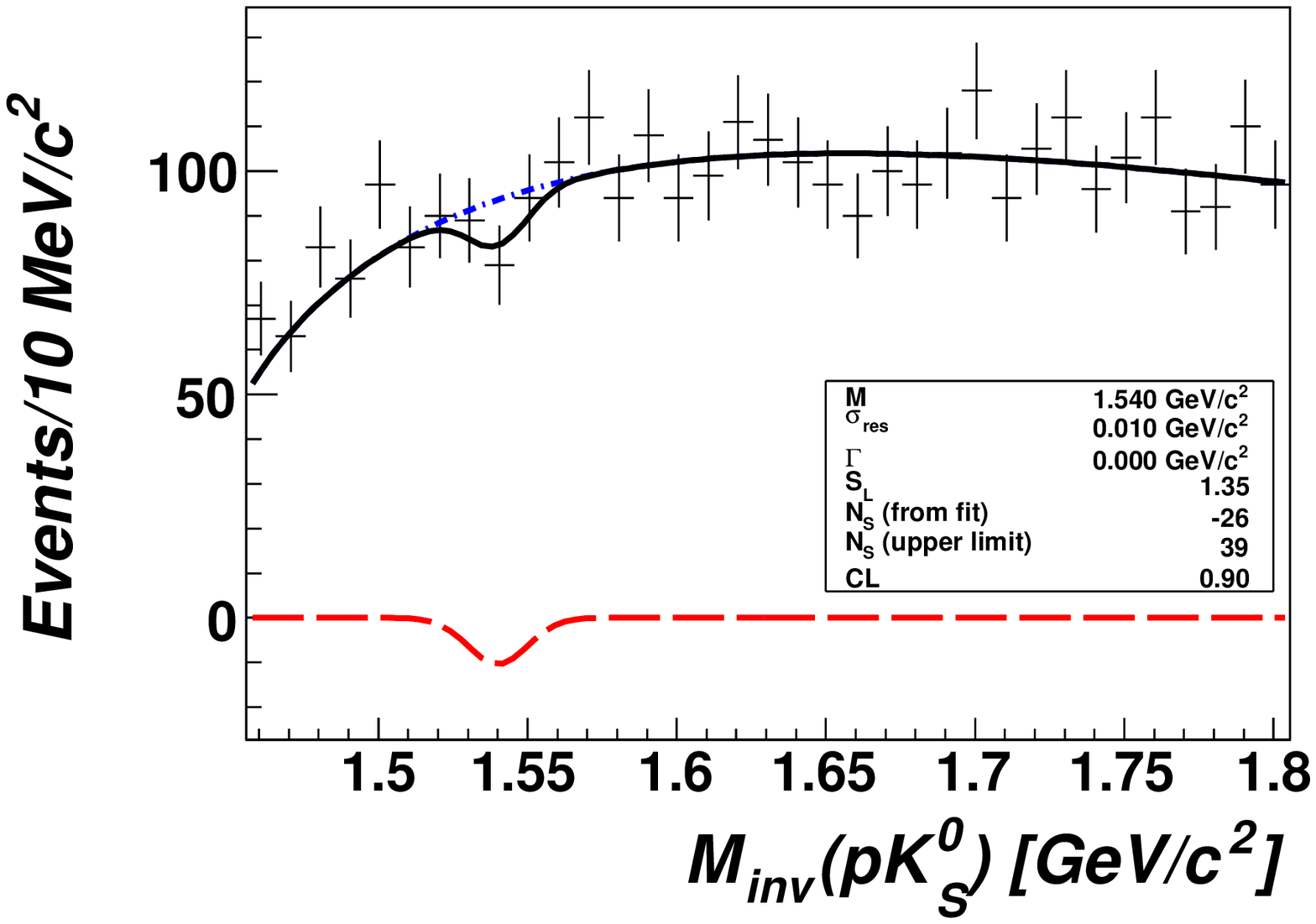,width=0.45\textwidth}\\
\end{tabular}
\caption{\label{fig:openbox-xf3} Invariant mass distributions of pairs of a
positively charged track (assumed to be a proton) and a $\ko$ in the data,
for $-0.3<x_F<0$. (Left) No proton identification; (Right) Optimized proton
identification.
The curves represent the predicted background and the amount of $\Theta^+$
signal maximizing $S_L$ (see Sec. 5.3).} 
\end{center}
\end{figure*}

\begin{figure*}[htb]
\begin{center}
\begin{tabular}{cc}
\epsfig{file=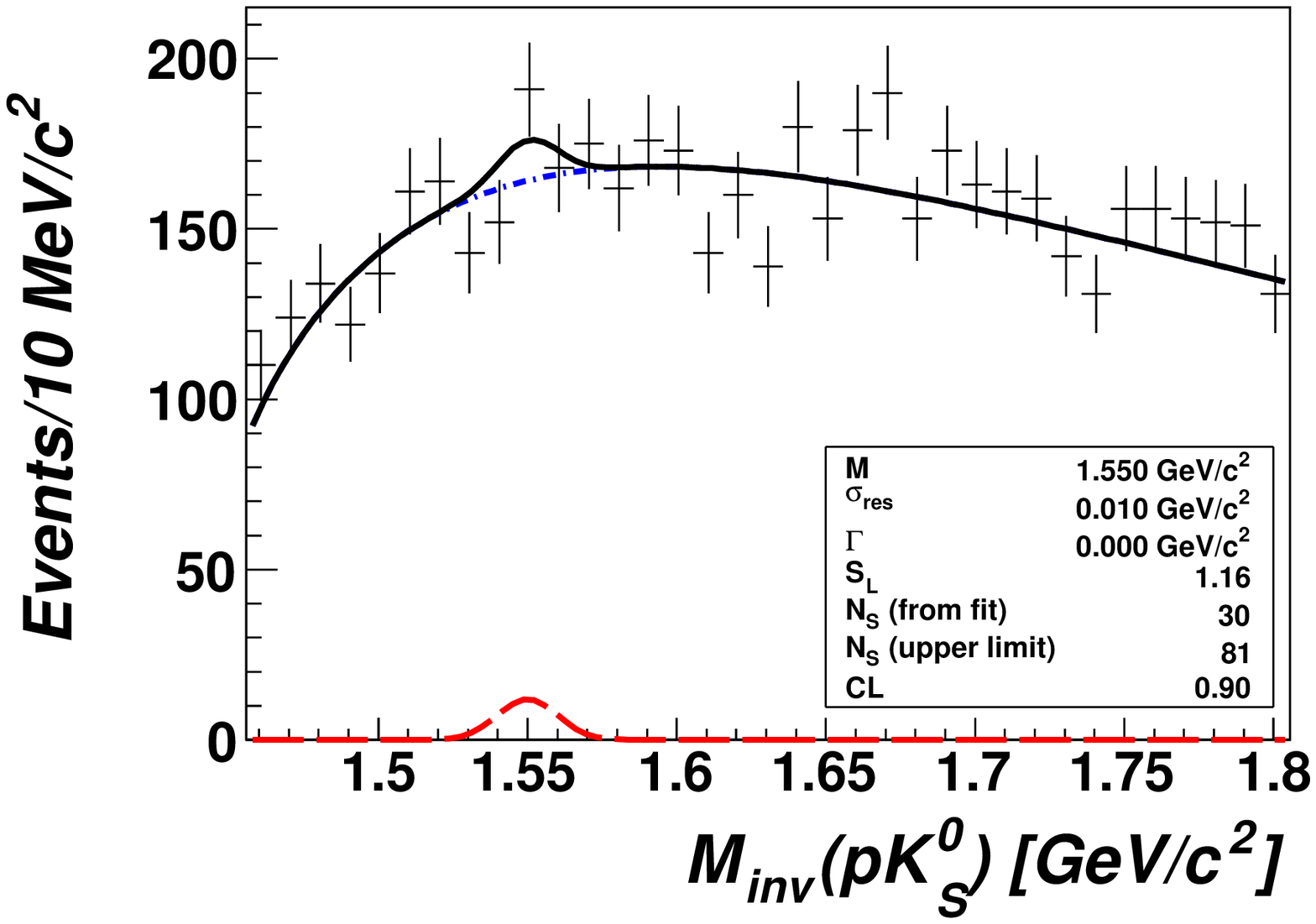,width=0.45\textwidth}&
\epsfig{file=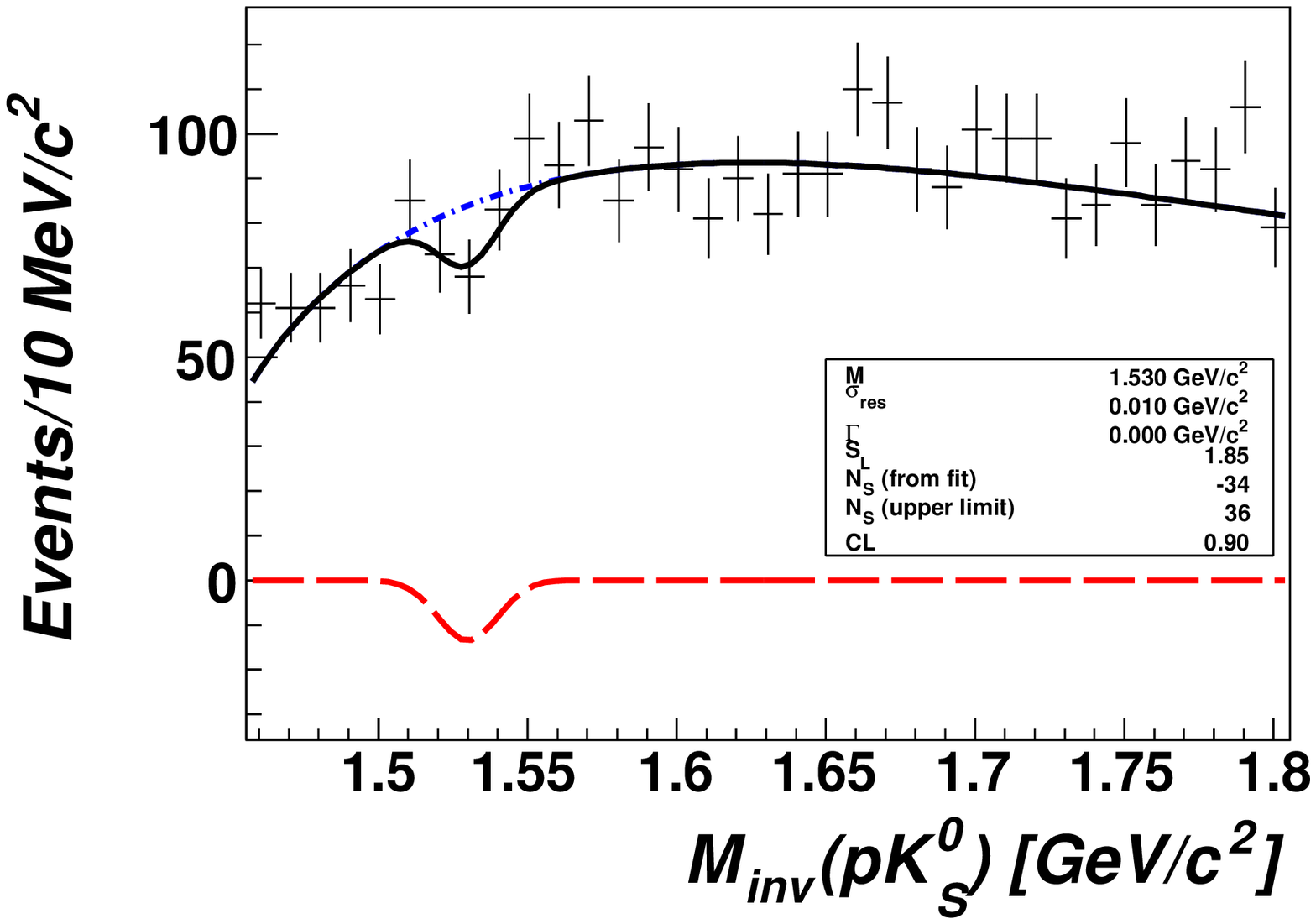,width=0.45\textwidth}\\
\end{tabular}
\caption{\label{fig:openbox-xf4} Invariant mass distributions of pairs
of a positively charged track (assumed to be a proton) and a $\ko$ in the data,
for $0<x_F<0.4$. (Left) No proton identification; (Right) Optimized proton
identification.
The curves represent the predicted background and the amount of $\Theta^+$
signal maximizing $S_L$ (see Sec. 5.3).} 
\end{center}
\end{figure*}

\begin{figure*}[htb]
\begin{center}
\begin{tabular}{cc}
\epsfig{file=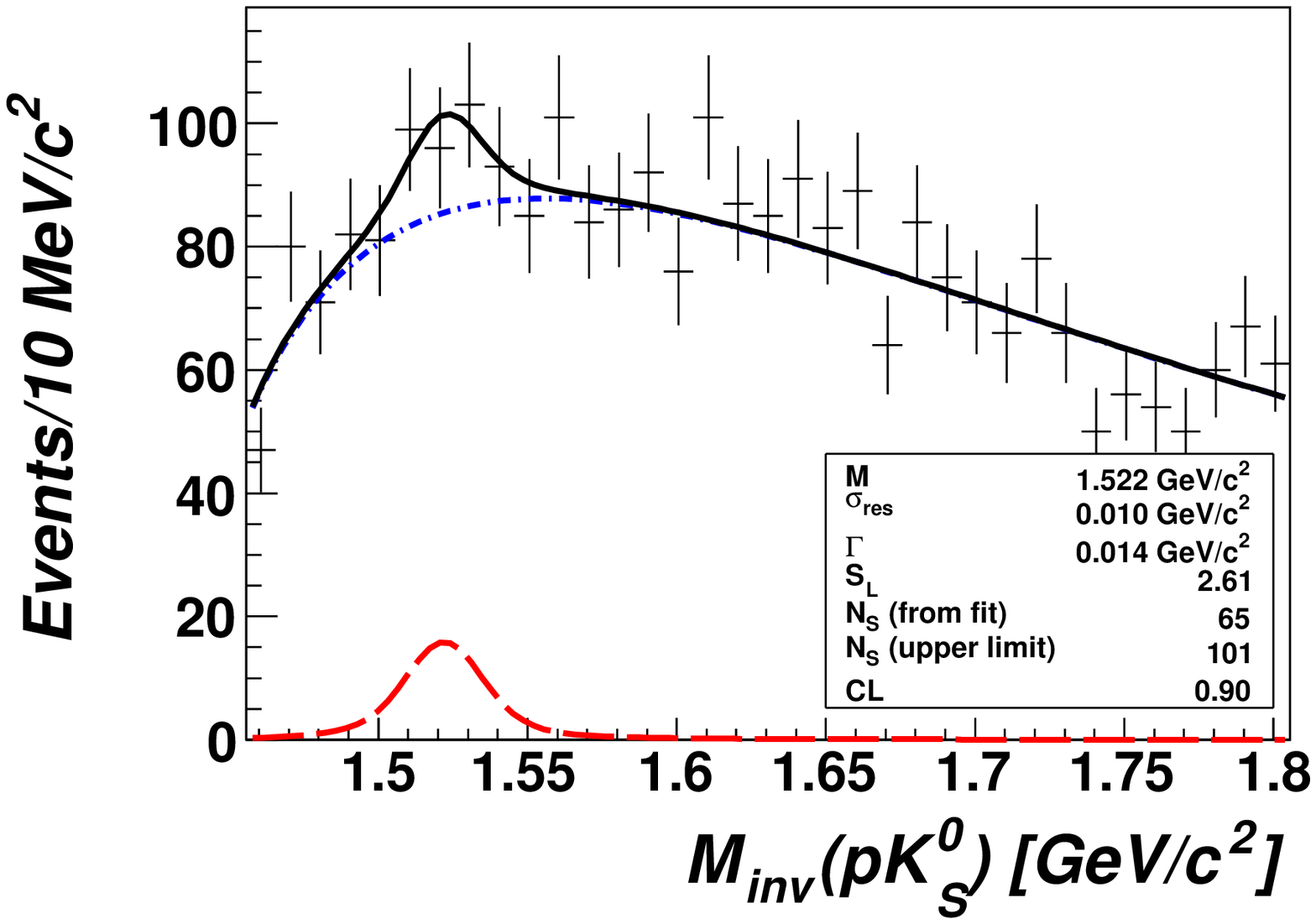,width=0.45\textwidth}&
\epsfig{file=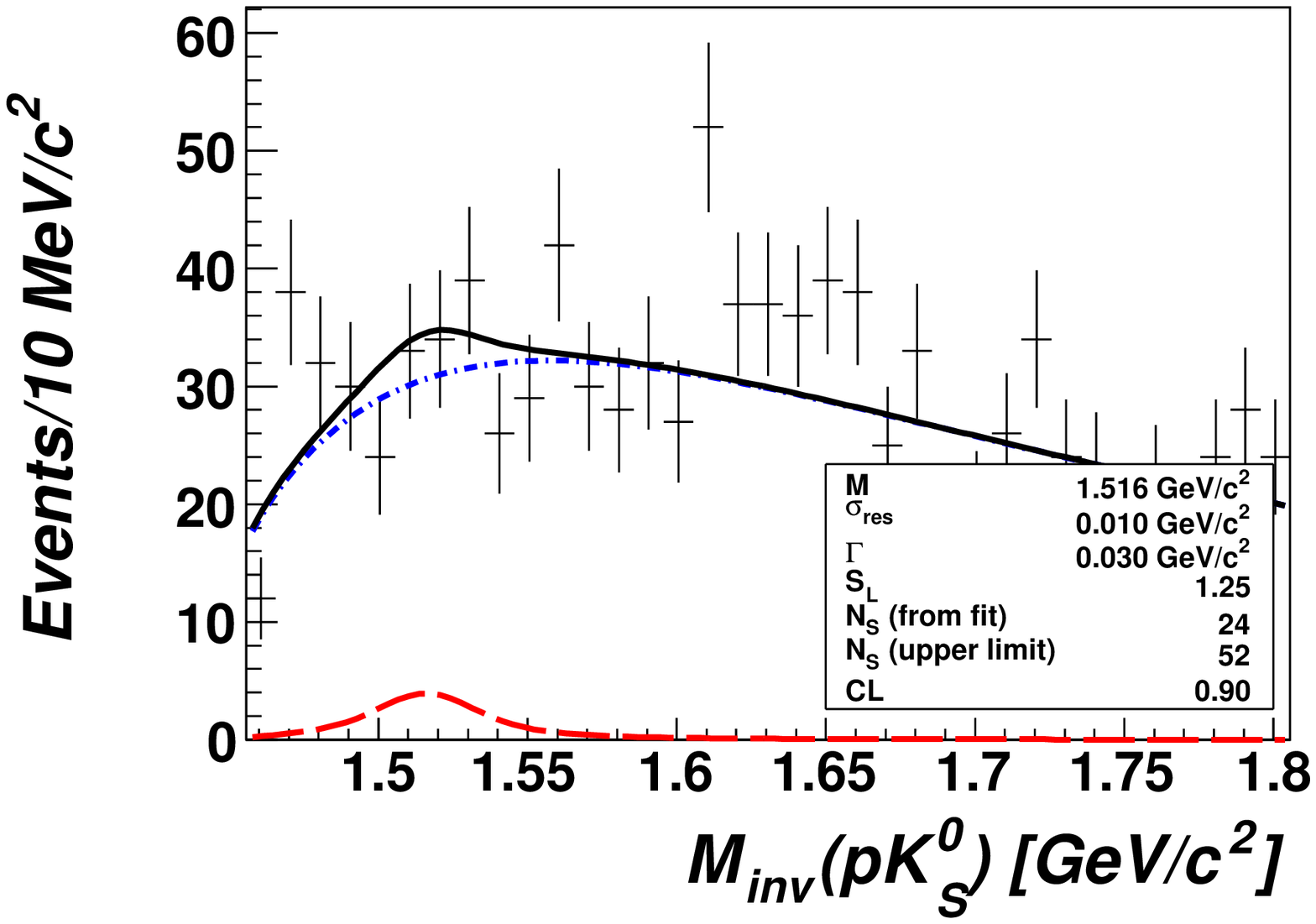,width=0.45\textwidth}\\
\end{tabular}
\caption{\label{fig:openbox-xf5} Invariant mass distributions of pairs
of a positively charged track (assumed to be a proton) and a $\ko$ in the data,
for $0.4<x_F<1$. (Left) No proton identification; (Right) Optimized proton
identification.
The curves represent the predicted background and the amount of $\Theta^+$
signal maximizing $S_L$ (see Sec. 5.3).} 
\end{center}
\end{figure*}

\begin{figure*}[htb]
\begin{center}
\begin{tabular}{cc}
\epsfig{file=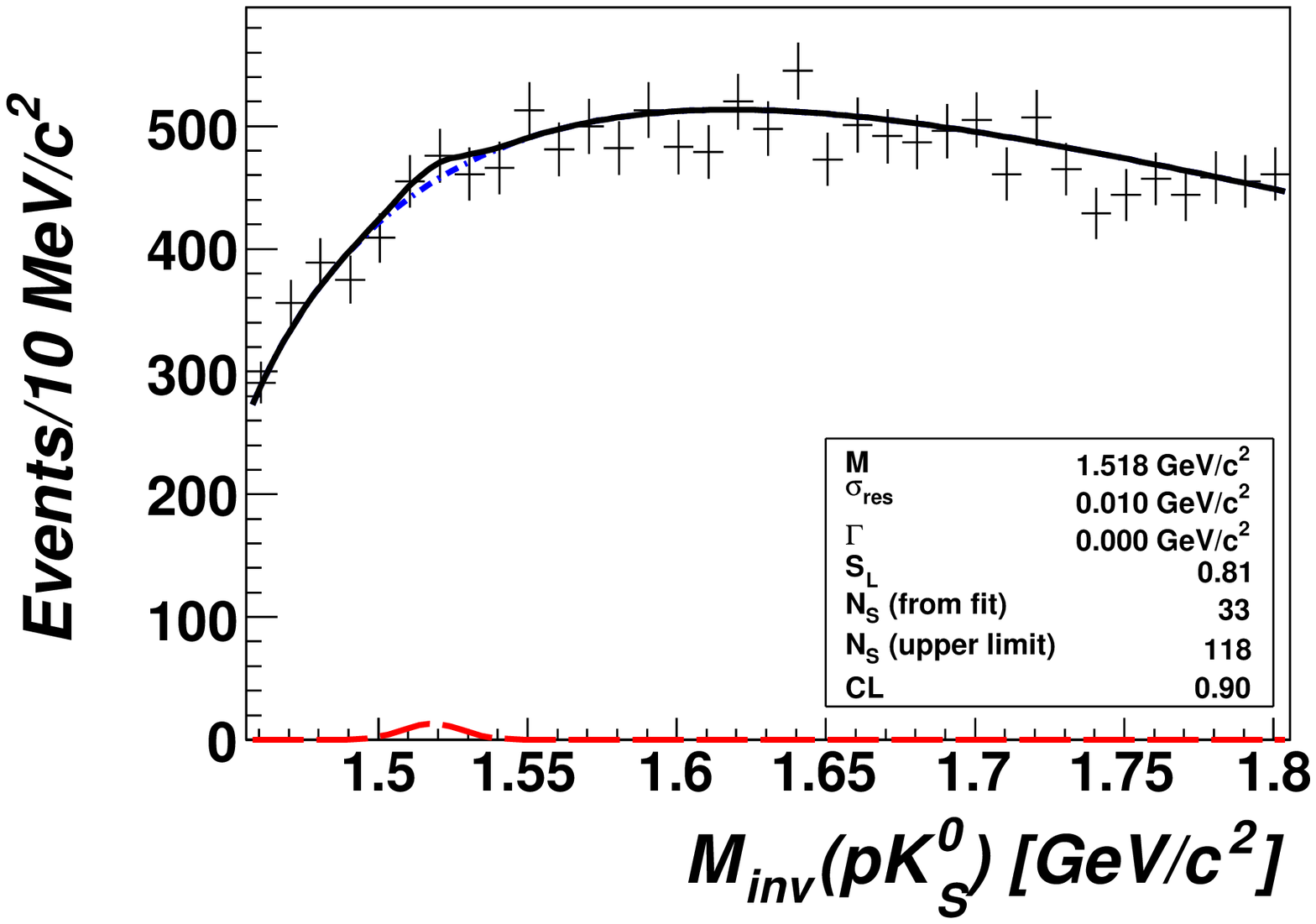,width=0.45\textwidth}&
\epsfig{file=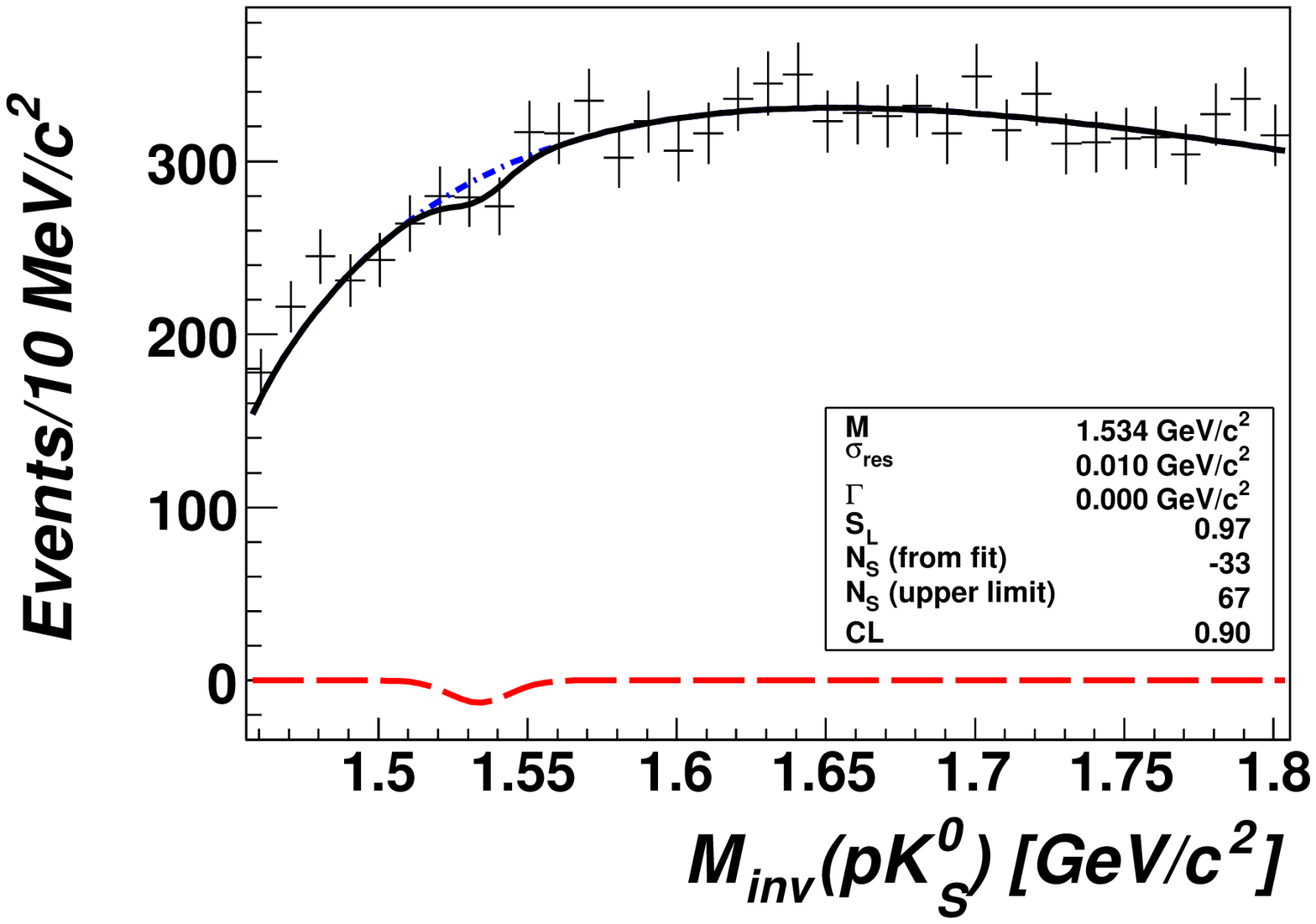,width=0.45\textwidth}\\
\end{tabular}
\caption{\label{fig:openbox-xfall} Invariant mass distributions of
pairs of a positively charged track (assumed to be a proton) and a $\ko$ in the
data, for $-1<x_F<1$. (Left) No proton identification; (Right) Optimized proton
identification.
The curves represent the predicted background and the amount of $\Theta^+$
signal maximizing $S_L$ (see Sec. 5.3).} 
\end{center}
\end{figure*}

We split the data into five $x_F$ intervals:\\
$[-1,-0.6)$, $(-0.6,-0.3)$, $(-0.3,0)$, $(0,0.4)$, $(0.4,1]$.
In each interval we optimize the proton identification cuts as described
in Sec.~\ref{sec:proton_id}, and estimate a possible signal in the region
$1510<M<1550$ MeV$/c^2$ as described in Sec.~\ref{sec:stat_analysis}.
Figs.~\ref{fig:openbox-xf1}-\ref{fig:openbox-xf5} display the results.
From these plots we conclude that we observe no evidence for the $\Theta^+$
state in any $x_F$ interval.
In Fig.~\ref{fig:openbox-xfall} we display the invariant mass distributions of
combinations of a positively charged track (assumed to be a proton) and a 
$\ko$ for the two cases of no proton identification and optimum proton
identification, for $-1<x_F<1$. 
%
Table~\ref{tab:upper_limits} summarizes the results and provides also the upper
limits at 90\% confidence level (CL) on the number of $\Theta^+$s candidates ($N_s^{up}$) and on
the production rate $R^{up}$ for both cases. The calculation of the
upper limits for the production rate include corrections for inefficiencies,
including the lack of detection of $K^0_L$ mesons, 
and take into account the $\kodecay$ branching ratio. The results are presented
for each bin of $x_F$, and also integrated over $x_F$. 
\begin{table*}[htb]
\begin{center}
     \begin{tabular}{||c|c|c|c|c|c|c||}
 \hline  \hline
     $x_F$ interval &  $[-1,-0.6)$ & $(-0.6,-0.3)$ &  $(-0.3,0)$ &  $(0,0.4)$ & $(0.4,1]$ & all \\
  \hline  \hline
    {\bf no ID}      &&&&& \\
 \hline
      $N_s$ (fit)	&  18       &26	        &35	   &30      &65          &77  \\
 \hline
      SL	& 1.96      &1.49       &1.01     &1.18	      &2.61	  &1.82 \\
 \hline
      $N_s^{up}$	 &41	   &61      &88	         &81	  &101	          &161   \\
 \hline
      $R^{up}$ & \it 3.84 & \it 2.18 & \it 1.74 & \it 1.37 & \it 0.83 & { \it 4.36}\\
 \hline
      \hline
      {\bf optimal ID} &&&&&&\\
 \hline
      $N_s$ (fit)	&  12     &   29       &   -26         &   -34       &  24             & -33  \\
 \hline
      SL	& 1.38      &   1.72   &   1.35    &   1.85    &    1.25     &  0.97   \\
 \hline
      $N_s^{up}$ &   28	   &     68   &     39     &    36    &      52     &   67   \\
 \hline
      $R^{up}$ & \it 2.80 & \it 2.60 & \it 0.84 & \it 0.79 & \it 1.00 & { \it 2.13} \\
 \hline  \hline
    \end{tabular}
\caption{\label{tab:upper_limits} Upper limits (90\% CL) on the number of
$\Theta^+$ candidates ($N_s^{up}$) and on the $\Theta^+$ production
 rate ($R^{up}$, in units of events per $10^3$ interactions) for the
 case of no proton identification and with optimal proton identification.}
\end{center}
\end{table*}
Fig.~\ref{fig:upper-limits} displays the sensitivity and upper limits (90\% CL)
for the $\Theta^+$ production rate as a function of $x_F$. The upper limits
are given as five curves, each corresponding to a fixed $\Theta^+$ mass,
obtained by varying both the number of signal events and the
$\Theta^+$ width to maximize $S_L$ as outlined in
Sec.~\ref{sec:stat_analysis}. 

\begin{figure*}[htb]
\begin{center}
\begin{tabular}{cc}
\epsfig{file=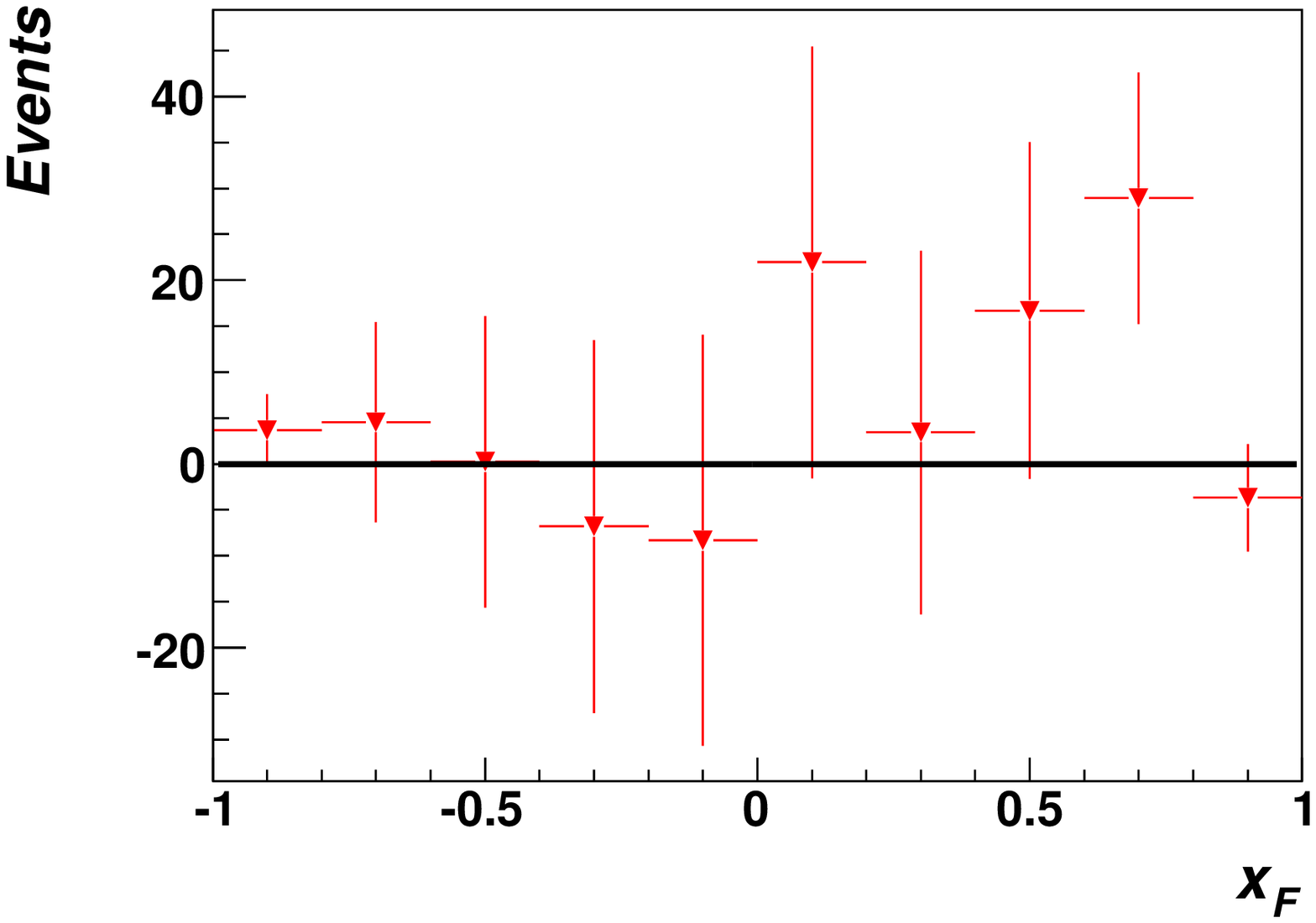,width=0.45\linewidth}
\epsfig{file=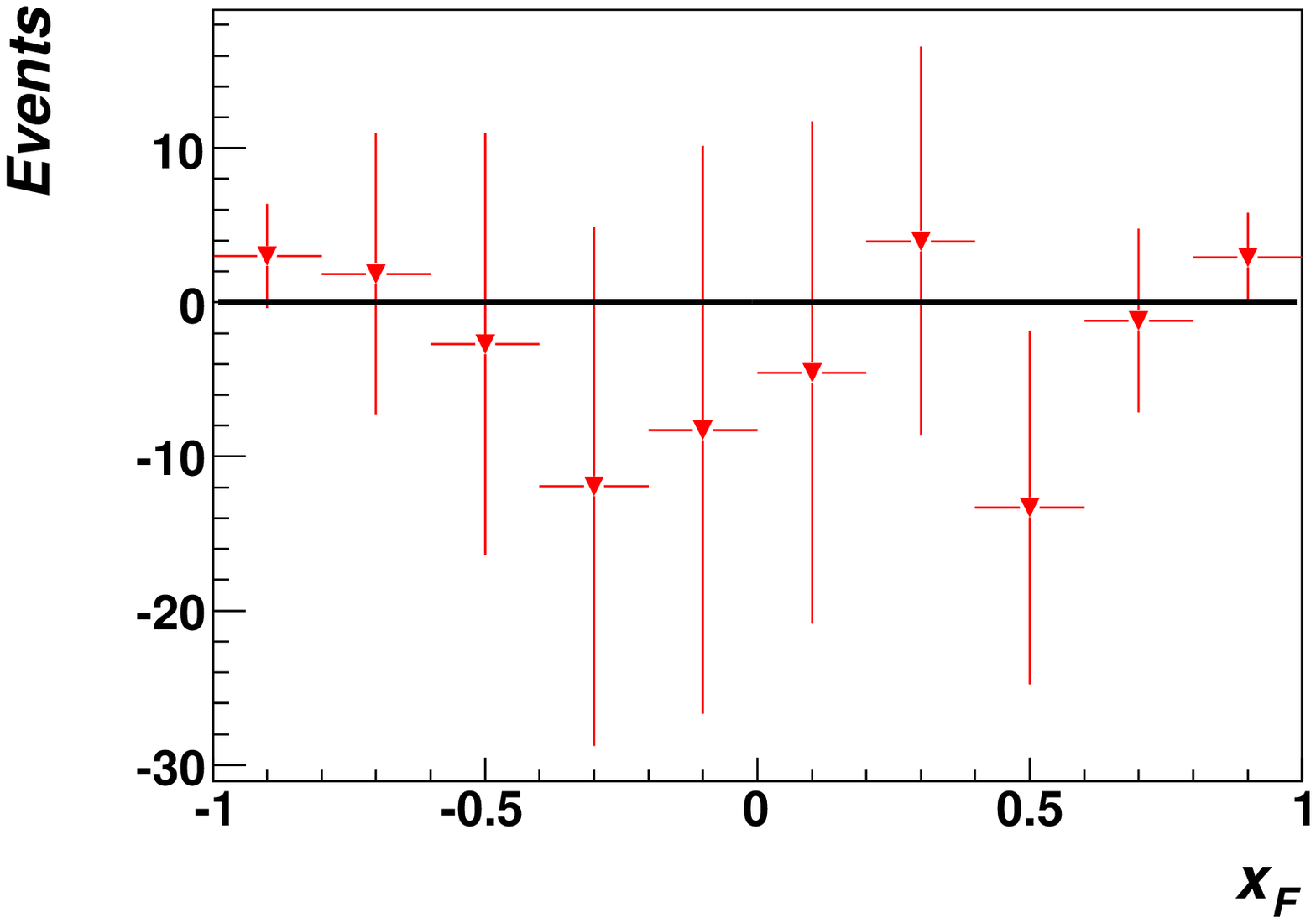,width=0.45\linewidth}
\end{tabular}
\caption{\label{fig:openbox-xf-distr} $x_F$ distribution of a potential
 $\Theta^+$ signal in the data. (Left) No proton identification; 
(Right) Optimized proton identification.}
\end{center}
\end{figure*}

We also measure the $x_F$ distribution of a potential $\Theta^+$ state
as follows. We build the $x_F$ distributions in two side-bands,
$1460<M<1500$ MeV$/c^2$ and $1580<M<1600$ MeV$/c^2$. We then normalize
the average of these two distributions to the expected number of background
events in the ``signal`` region ($1510<M<1550$ MeV$/c^2$), and subtract it
from the $x_F$ distribution of the data in the ``signal`` region. The
result can be considered as the $x_F$ distribution of the signal, and
could shed a light on the $\Theta^+$ production mechanism.
Fig.~\ref{fig:openbox-xf-distr} displays the result with no proton
identification and with optimal proton identification. We observe no
statistically significant accumulation of events at any $x_F$ value.


\section{\label{sec:conclusion} Conclusions}

We have performed a blind search for the $\Theta^+$ exotic baryon in the
$\Theta^+\to p+\ko$ decay mode in the NOMAD $\nu_\mu N$ data. We have
built a robust background estimation procedure which has been tested against
various known cases like $\lamdecay$, $\kodecay$ and $\Kdecay$. In all
cases good agreement between data and estimated background has been found.
Good agreement has also been found between the invariant mass ($M$)
distribution of $\ko$--proton pairs in the data and the estimated background
in the whole mass region excluding the $\Theta^+$ signal region. We have
developed proton identification tools based on the discrimination power of
three sub-detectors, and we have tuned the proton identification criteria by
maximizing the sensitivity to the expected signal in five $x_F$ intervals
independently. We have checked this approach for $\lamdecay$ and found
that this procedure indeed maximizes the signal significance in both MC and
data. Finally, we have ``opened the box``, i.e. examined the $\Theta^+$ signal
in the data and found good agreement between the data and the background for
the whole $M$ region, including the ``signal`` region, in each $x_F$ interval.
We observe no evidence, therefore, for any $\Theta^+$ signal in the 
$\Theta^+\to p+\ko$ decay channel in the NOMAD $\nu_\mu N$ data. We give
an upper limit  at 90\% CL on $\Theta^+$ production rate of $2.13\cdot10^{-3}$
events per neutrino interaction at $M=1530$ MeV$/c^2$ after integrating
over $x_F$.

\begin{figure}[htb]
\begin{center}
\epsfig{file=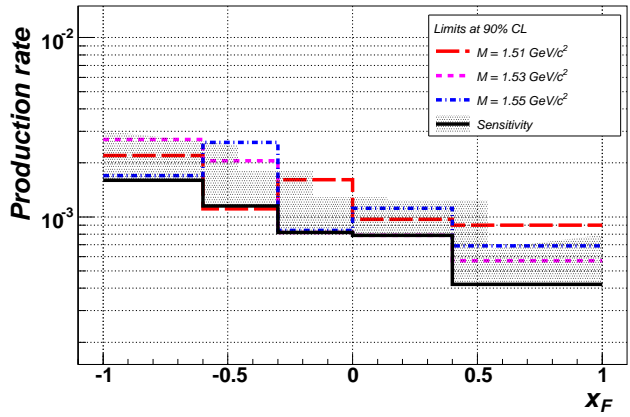,width=\linewidth}
\caption{\label{fig:upper-limits} Sensitivity and upper limits at 90\% CL for
$\Theta^+$ production rates as a function of $x_F$, for $\Theta^+$ masses
of 1510, 1530, 1550 MeV$/c^2$.}
\end{center}
\end{figure}

It is interesting to compare this result with the recent analysis of old bubble
chamber neutrino experiments which provide an estimation of the $\Theta^+$
production rate as large as $\sim 10^{-3}$~ events per neutrino interaction\cite{Bebc}.
As shown in Fig.\ref{fig:upper-limits}, for a large fraction of the $x_F$
range, except in the region $x_F\to -1$, such a value is excluded.
Unfortunately, ref.~\cite{Bebc} does not provide information on the $x_F$
region in which a $\Theta^+$ signal was observed. Furthermore, in
ref.~\cite{Bebc} we find no information that the background estimation
procedure took into account the effects mentioned in Sec.~\ref{sec:background},
which can result in an underestimation of the background and thus in an
overestimation of both the signal significance and the production rate.

\begin{figure*}[htb]
 \begin{center}
  \begin{tabular}{cc}
   \epsfig{file=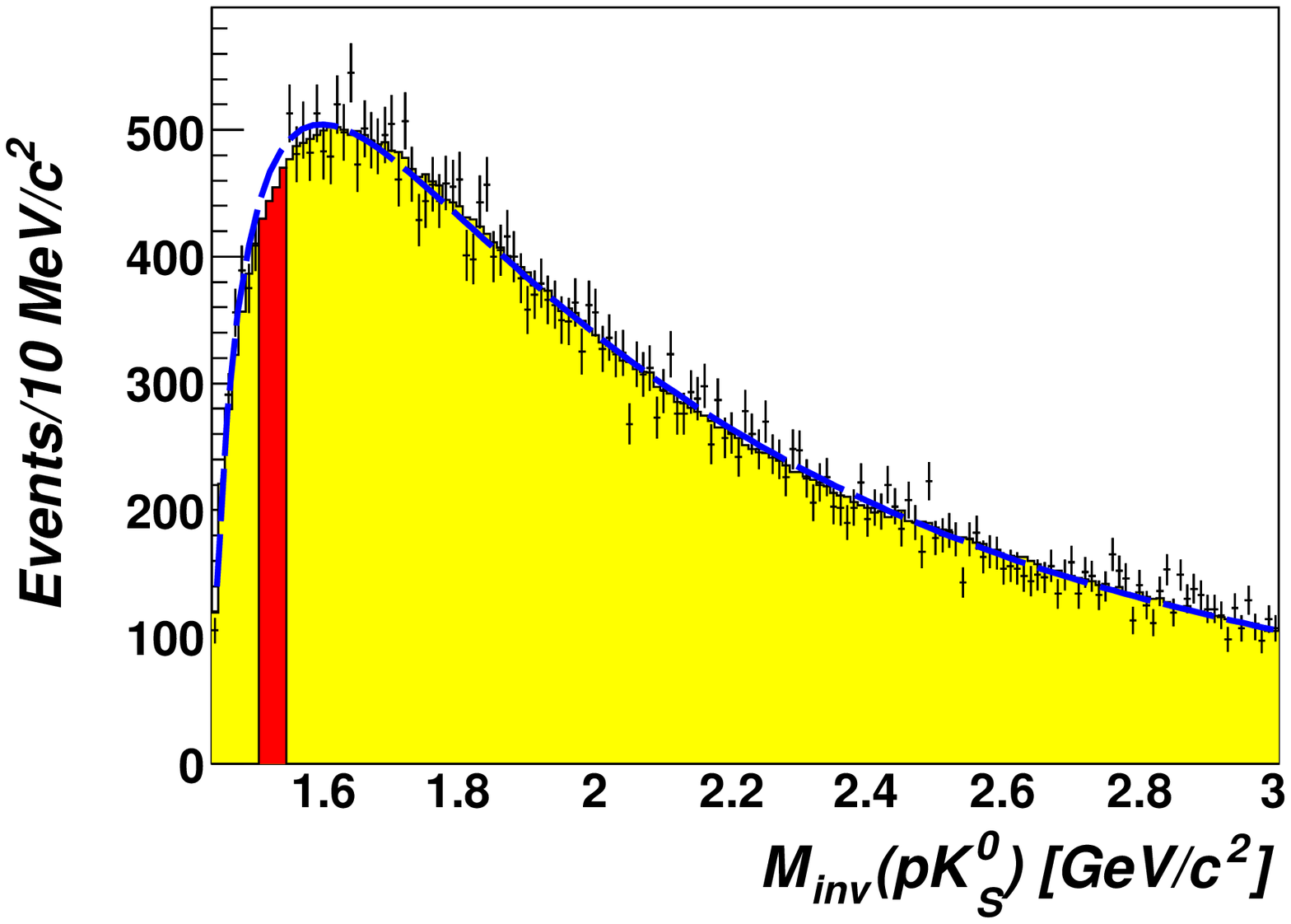,width=0.4\linewidth}&
   \epsfig{file=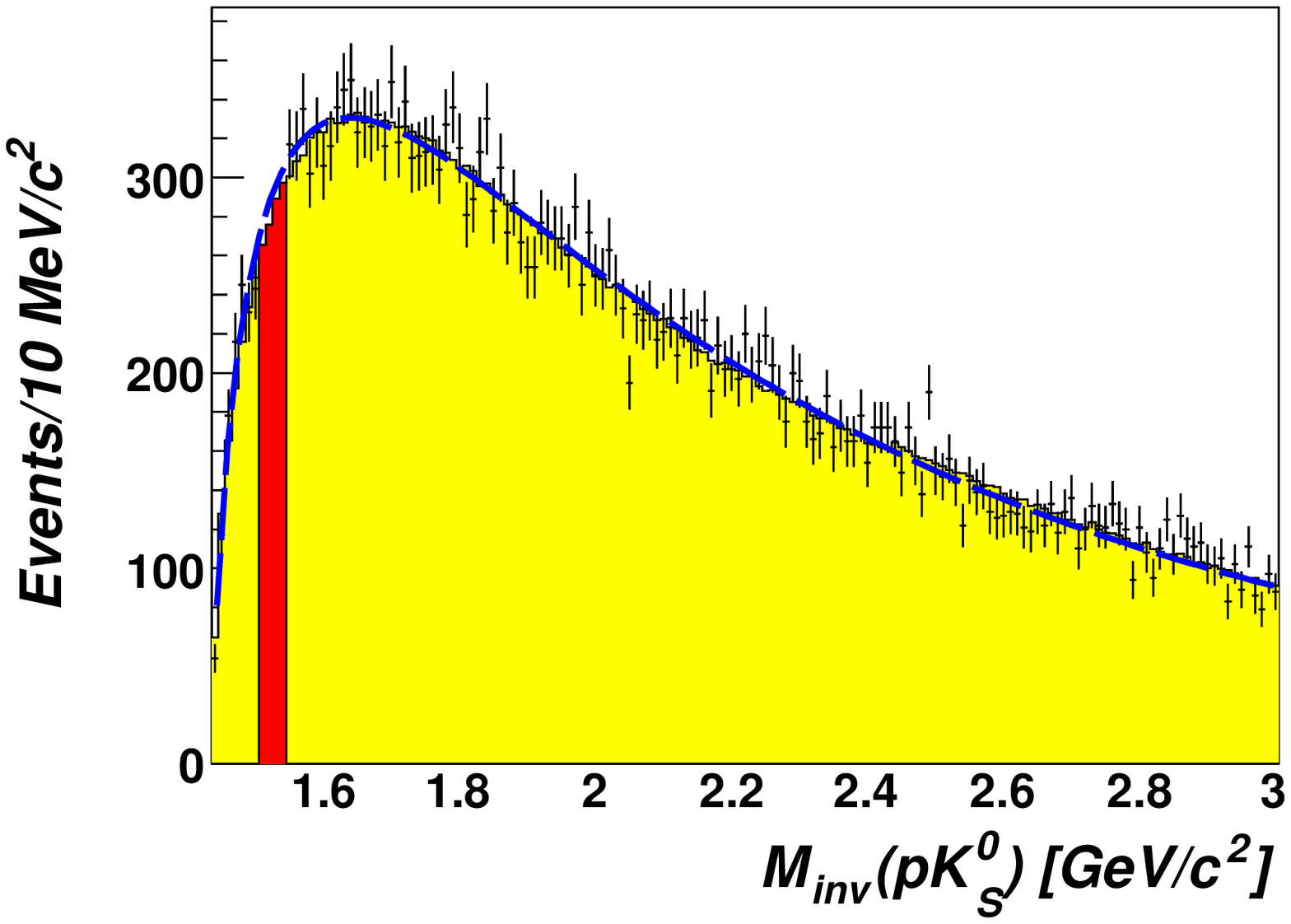,width=0.4\linewidth}\\
  \end{tabular}
  \caption{\label{fig:data_vs_background_blind} Invariant mass distribution of pairs of one positively charged track (assumed to be a proton) and a $\ko$ for the data (points with error bars) and for the fake pair background (shadowed area). (Left) Proton identification has not been used. (Right) ``Optimal`` proton identification. Data in the ``signal`` region ($1510< M < 1550$ MeV$/c^2$) are not shown. Dashed curve is fit of the data by a polynomial excluding the ``signal`` region.}
\end{center}
\end{figure*}

Preliminary NOMAD results from searches for the exotic $\Theta^+$ baryon reported earlier ~\cite{Camilleri:2005hc}, quoting a hint for a signal with a statistical significance of 4.3 $\sigma$, suffered from an incorrect background estimation, which did not take into account the effects
mentioned in Sec.~\ref{sec:background}. 
The results reported in  ~\cite{Camilleri:2005hc} are obtained on a smaller sample of the NOMAD data. The positives of that sample were subjected for a cleaner proton identification which yielded an increase of the purity of the protons sample from 23\% to 51.5\% with about factor six lost of the statistics. The difference in shapes of $\ko p$ invariant mass distributions reported in  ~\cite{Camilleri:2005hc} and in Figs.~\ref{fig:data_vs_background_blind} is due to an additional requirement imposed in ~\cite{Camilleri:2005hc} on energy of protons to be larger than that of $\ko$. 

\begin{acknowledgement}
We gratefully acknowledge  the CERN SPS accelerator and beam-line staff
for the magnificent performance of the neutrino beam. The experiment was 
supported by the following funding agencies: Australian Research Council 
(ARC) and Department of Education, Science, and Training (DEST), Australia;
Institut National de Physique Nucl\'eaire et Physique des Particules (IN2P3), 
Commissariat \`a l'Energie Atomique (CEA), France; Bundesministerium 
f\``ur Bildung und Forschung (BMBF, contract 05 6DO52), Germany; Istituto 
Nazionale di Fisica Nucleare (INFN), Italy; Joint Institute for Nuclear 
Research and Institute for Nuclear Research of the Russian Academy of 
Sciences, Russia; Fonds National Suisse de la Recherche Scientifique, 
Switzerland; Department of Energy, National Science Foundation 
(grant PHY-9526278), the Sloan and the Cottrell Foundations, USA. 

We also thank A.~Asratyan and A.~Dolgolenko for valuable discussions.
\end{acknowledgement}

\bibliographystyle{unsrt}
\bibliography{references}
\cleardoublepage

\clearpage

\clearpage

\clearpage

\clearpage

\clearpage

\clearpage

\clearpage

\clearpage

\clearpage

\clearpage

\clearpage

\clearpage

\clearpage

\clearpage

\clearpage

\end{document}